\documentclass[onecolumn]{aa}

\newcommand{\gbm}{\textit{Fermi}/GBM}

\newcommand{\Ep}{$E_{p,z}$}
\newcommand{\Liso}{$L_{iso}$}
\newcommand{\ratio}{$\nu_m/\nu_c$}

\usepackage{txfonts}
\usepackage{graphicx}
\usepackage{hyperref}
\usepackage{amssymb}
\usepackage{comment}
\usepackage{amsmath}
\usepackage{soul}
\usepackage{booktabs}
\usepackage{url}
\usepackage{multirow}
\usepackage{tcolorbox}
\usepackage{lineno}
\usepackage{mathtools}
\usepackage{caption}
\usepackage{subcaption}
\usepackage[thinc]{esdiff}
\usepackage{xtab,booktabs}

\begin{document} 

\title{Gamma-ray burst spectral-luminosity correlations in the synchrotron scenario}
\titlerunning{Gamma-ray burst spectral-luminosity correlations in the synchrotron scenario}

\author{Alessio Mei \inst{1,2}\thanks{\email{alessio.mei@gssi.it}}
       \and
       Gor Oganesyan\inst{1,2}
       \and 
       Samanta Macera\inst{1,2}   
}
\institute{ Gran Sasso Science Institute (GSSI), Via F. Crispi 7, 67100 L'Aquila, Italy\\
    \and
    INFN—Laboratori Nazionali del Gran Sasso, I-67100, L’Aquila (AQ), Italy\\
}

%\date{Received September 15, 1996; accepted March 16, 1997}
\date{}

\abstract
{For over two decades, gamma-ray burst (GRB) prompt emission spectra were modelled with smoothly-broken power laws (Band function), and a positive and tight correlation between the spectral rest-frame peak energy \Ep\ and the total isotropic-equivalent luminosity \Liso\ was found, constituting the so-called Yonetoku relation. However, more recent studies show that many prompt emission spectra are well described by the synchrotron radiation model, hence significantly deviating from the Band function.}
{In this work, we test the impact of a more suited spectral model such as an idealized synchrotron spectrum from non-thermal electrons on the Yonetoku relation and its connection with physical parameters.}
{We select GRBs with measured redshift observed by \gbm\ together with high energy observations (>30 MeV), and perform spectral analysis dividing them in two samples: the \textit{single-bin} sample, using the light curve peak spectrum of each GRB, and the \textit{multiple-bins} sample, where we explore the whole duration of 13 bright bursts with time-resolved spectral analysis.}
{We observed that the \Ep\ of synchrotron spectra in fast-cooling regime (\ratio\ $\gg1$) is generally larger than the one provided by the Band function. For this reason, we do not find any \Ep$-$\Liso\ correlation in our samples except for the GRBs in an intermediate-cooling regime ($1<$\ratio$<3$), namely where peak and break energies are very close. We instead find in both our samples a new tight correlation between the rest-frame cooling frequency $\nu_{c,z}$ and \Liso: $\nu_{c,z} \propto L_{iso}^{(0.53 \pm 0.06)}$.}
{These results suggest that, assuming that prompt emission spectra are produced by synchrotron radiation, the physical relation is between $\nu_{c,z}$ and $L_{iso}$. The fit of the Band function to an intrinsic synchrotron spectrum returns peak energy values $E_{p,z}^{Band} \sim \nu_{c,z}$. This may explain why the systematic interpretation of prompt spectra through the Band function returns the $E_{p,z}-L_{iso}$ relation.}


\maketitle

\section{Introduction} \label{sec:intro}

Cosmic explosions triggered by collapses of massive stars \citep{Woosley1993,Woosley2006} or binary neutron star mergers \citep{Eichler1989,Narayan&Paczynski1992,Berger2014, Abbott2017} produce gamma-ray bursts (GRBs), the most luminous transient phenomena in the Universe \citep{Paczynski1986}.\\
Thanks to several decades of observations, GRBs revealed to be associated to highly collimated ultra-relativistic jets launched as an aftermath of their progenitor's explosion (see \citealt{Salafia&Ghirlanda2022} for a review).\\
The very fast prompt emission variability (e.g. \citealt{Bhat2012}) clearly suggests an emission site internal to the jet \citep{Sari&Piran1997}. A natural explanation for the jet kinetic energy dissipation is through internal shocks \citep{Rees&Meszaros1994}. However, sub-photospheric dissipation \citep{Rees&Meszaros2005, Pe'er2008} or magnetic reconnection processes \citep{Drenkhahn&Spruit2002, Zhang&Yan2011} are equally able to explain the GRB radiative output.\\
The early emission observed soon after the explosion, during the so-called prompt emission phase, shows a broad and peaked spectral energy density (SED), possibly associated with non-thermal radiative processes. 
Historically, prompt emission spectra were modelled by the phenomenological Band function \citep{Band1993}, namely two power laws smoothly connected around a peak at energy $E_p$. 
The systematic fit of this model to large samples of GRB spectra revealed that in some cases the low energy photon index $\alpha$ was consistent with the one predicted by synchrotron emission in a fast-cooling regime, whereas some other GRBs exhibited very hard $\alpha$, pointing towards quasi-thermal processes \citep{Ghisellini&Celotti1999, Lazzati2000, Meszaros&Rees2000}. The vast majority of the GRBs showed an average behaviour, with spectra too hard to be produced by synchrotron processes, but too soft to advocate for thermal origins \citep{Preece1998, Kaneko2006, Gruber2014}.\\
This tension was partially mitigated when more complex models were employed and low energy spectral breaks were identified in some GRB spectra at energies $E_b$. It was shown
% While some authors highlight a second thermal superimposed to a main non-thermal one at low energies \citep{Ghirlanda2003,Ryde2005,Guiriec2013, Guiriec2015, Burgess2014, Yu2015}
that a single synchrotron spectrum in marginally fast-cooling regime \citep{Oganesyan2017, Oganesyan2018, Ravasio2019} or slow-cooling regime \citep{Zhang2009,Burgess2020} are able to account for the entire prompt emission spectrum, from the optical \citep{Oganesyan2019} up to the GeV band (Macera et al., article in preparation). Despite the synchrotron model seems to be a viable explanation for most of the considered spectra, GRBs show diverse spectral and temporal behaviours, preventing a convincing unified interpretation.\\
A feature that appears to be shared among GRBs is that their prompt spectrum is harder when the GRB is brighter and more energetic. In fact, many sources show that the rest-frame spectral peak energy $E_{p,z}$ is positively and tightly correlated with the isotropic-equivalent energy $E_{iso}$ \citep{Amati2002} and luminosity $L_{iso}$ \citep{Yonetoku2004}. These two relations span several decades in both energies and luminosities, and despite being affected by selection and observational biases \citep{Band&Preece2005, Nakar&Piran2005, Butler2007, Shahmoradi&Nemiroff2011}, they seem to hold for most of the GRB class \citep{Ghirlanda2008, Nava2008}, also at different redshifts \citep{Nava2012}.\\
In particular, the $E_{p,z}-E_{iso}$ relation (better known as the "Amati relation") was initially found for long GRBs, with durations $T_{90}>2\ {\rm s}$. Short GRBs ($T_{90}<2$ s) exhibit a similar trend but occupy a parallel track in the $E_{p,z}-E_{iso}$ plane \citep{Ghirlanda2009}, showing a systematically lower $E_{iso}$ for the same $E_{p,z}$. On the other hand, both long and short GRBs follow the same $E_{p,z}-L_{iso}$ relation (better known as the "Yonetoku relation", \citealt{Yonetoku2010}). After their first discoveries, many outliers to these relations were found, however both the Amati and Yonetoku relations still hold when multiple time-bins from the same bright GRB are considered through time-resolved spectral fits \citep{Ghirlanda2010, Ghirlanda2011a, Ghirlanda2011b} .\\
At odds with the Amati relation, which is obtained comparing quantities averaged across the whole burst duration, the Yonetoku relation is found considering \Liso\ (but also \Ep, e.g. \citealt{Tsvetkova2017}) estimates only during the brightest pulse, i.e. at the peak of the light curve. This makes the Yonetoku relation inherently connected with the physics taking place in single GRB pulses, and it appears particularly suited to inquire prompt emission physics in depth.\\
For more than two decades, the observations carried out by gamma-ray instruments such as the Burst Alert Telescope (BAT, 15–150
keV) on board the Neil Gehrels Swift Observatory (\textit{Swift}), Gamma-ray Burst Monitor (GBM, 8 keV–40 MeV) on board the \textit{Fermi} satellite and  Konus-Wind (KW, $\sim20$ keV–20 MeV) further corroborated the validity of these correlations (e.g. \citealt{Nava2012, Tsvetkova2017, Minaev2020, Wang2024}).\\
Despite all these findings point towards a possible common emission mechanism occurring in GRB central engines, a clear physical interpretation of these correlations is still missing (see \citealt{Parsotan&Ito2022} for a review).\\
In this work, we carry out spectral fits of GRB samples detected by \gbm\ using not only the phenomenological Band function, but also a physical synchrotron model. Our goals are (i) to test whether using a physical model allows to derive the \Ep$-$ \Liso\ (Yonetoku) relation and (ii) to explore any possible connection between the Yonetoku relation and physical synchrotron parameters such as the characteristic minimum frequency $\nu_m$ and the cooling frequencies $\nu_c$.\\
The paper is organized as follows: in Section \ref{sec:methods} we describe the sample selection and the spectral modelling; in  Section \ref{sec:results} we show the results of the spectral fit to the whole sample of GRBs (\textit{single-bin} sample) and the time-resolved analysis of a sub-sample of bright GRBs (\textit{multiple-bins} sample); in Section \ref{sec:discussion} we discuss the results, and finally in Section \ref{sec:conclusion} we summarize the conclusions. Throughout the paper we assume standard cosmology with $h = \Omega_\Lambda = 0.7$ and $\Omega_m= 0.3$. If not stated differently, errors are reported at $68\%$ confidence level. As a convention, the subscript $z$ refers to rest-frame parameters.\\

\section{Methods} \label{sec:methods}

\subsection{Sample selection}\label{sec:sample_selection}
We analyse GRBs observed by the \textit{Fermi} satellite before September 2023. We select the GRBs with a firm redshift estimate, as provided by the MPE GRB online catalog\footnote{\url{https://www.mpe.mpg.de/~jcg/grbgen.html}}. We consider only the GRBs that are bright enough during their main pulse to perform a robust spectral analysis in that temporal bin. Specifically, we include in our sample only the GRBs whose 1s peak photon flux in the energy range 50-300 keV $P\geq 3.5\ {\rm photons\ s^{-1}\ cm^{-2}}$, which corresponds to nearly 7 times the flux sensitivity of \gbm. We exclude GRB221009A, GRB230307A and GRB130427A, the three brightest GRBs, since their emission during the peak are affected by pile-up effects and their GBM data cannot be used for spectral analysis.\\
This leads to a total sample of 74 GRBs, with a spectroscopic redshift estimate for 70 of them. Three GRBs (GRB100816A, GRB200826A and GRB211211A) have a redshift estimate through host galaxy association while one GRB (GRB200829A) has a photometric redshift estimate.

\subsection{Data reduction}
We downloaded the \gbm\ data (8 keV - 40 MeV) of all the 74 GRBs in our sample from the Fermi GBM Burst catalogue\footnote{\url{https://heasarc.gsfc.nasa.gov/W3Browse/fermi/fermigbrst.html}}, and performed a standard data reduction using the Fermi science tool \textsc{GTBURST}\footnote{\url{https://fermi.gsfc.nasa.gov/ssc/data/analysis/scitools/gtburst.html}}. In particular, we consider data from the two NaI and one BGO detectors with best observational conditions (i.e. lowest viewing angles). In three GRBs (GRB121128A, GRB131231A, GRB140508A) the second best NaI was exhibiting artefacts in the low energy data ($<30$ keV) leading to spectral deviations with respect to the best NaI. For this reason, for these three GRBs we considered the best and third-best NaI, together with the best BGO detector.\\
The background analysis was performed through \textsc{GTBURST} by using, whenever the polynomial fit was converging, the background intervals reported by \textit{Fermi} collaboration \citep{vonKienlin2020}. In case this selection was leading to a non convergent background polynomial fit, larger custom background intervals were selected.\\
We selected the source interval to coincide with the peak of the GRB light curve. The time-bin is the one reported in the \texttt{bcat} file, a file containing basic burst parameters provided by Fermi collaboration for every GRB. The width of the time bin is $1.064$ s for long GRBs ($T_{90} > 2$ s) and $256$ ms for short GRBs ($T_{90} < 2$ s), in order to pinpoint their particularly short peak phase.\\
The outputs of the data reduction are the source, background and weighted response spectral files that are used for the spectral analysis through the Heasarc package \textsc{XSPEC}\footnote{\url{https://heasarc.gsfc.nasa.gov/xanadu/xspec/}} \citep{Arnaud1996}.\\
The addition of high energy ($>30$ MeV) data to the prompt emission proved to be crucial in determining spectral parameters otherwise inaccessible or difficult to constrain. For this reason, we decided to include in our dataset Large Area Telescope (LAT) Low Energy (LLE, 30-100 MeV) data whenever available. 
We retrieved the data from the Fermi LAT Low-Energy Events Catalog\footnote{\url{https://heasarc.gsfc.nasa.gov/W3Browse/fermi/fermille.html}}. 
LLE data are reduced using the same tools employed in the \gbm\ analysis, selecting the source in the same time-bin in order to perform a simultaneous joint time-resolved analysis during the brightest peak of the emission. In our current sample, only 22 GRB have available LLE data during their peak. For those GRBs we produce spectral files similarly to the GBM ones.

\subsection{Spectral fit routine}
After the data reduction, we model the spectra in our sample. For the fitting process we use \textsc{pyXSPEC} \footnote{\url{https://heasarc.gsfc.nasa.gov/xanadu/xspec/python/html/index.html}}, the Python interface to the \textsc{XSPEC} spectral-fitting program. We implemented the python-based software Bayesian X-ray Analysis \citep[BXA,][]{Buchner2016}, which allows for Bayesian parameter estimation and model comparison through nested-sampling algorithms in \textsc{pyXSPEC}. \\
We ignored the energy channels outside 8-900 keV for the NaI detectors, as well as the 30-40 keV band in order to avoid the Iodine K-edge line at 33.17 keV \citep{Meegan2009}. We selected the energy range 300 keV - 40 MeV and 30-100 MeV for BGO and LLE data, respectively.
For each dataset, consisting of GBM and LLE data (when available), we test two models: the phenomenological Band function (\texttt{grbm} in \textsc{XSPEC} notation, \citealt{Band1993}) and an idealised physical Synchrotron model.
We included the presence of a cross-calibration constant, \texttt{constant} in \textsc{XSPEC} notation, allowing for a $30\%$ variation for each dataset. We applied a Poisson-Gaussian statistics (\texttt{pgstat}) to GBM data and Cash statistics (\texttt{cstat}) to LLE data.\\
For each model we measure the logarithm of the 10 keV - 10 MeV bolometric flux in the source rest frame $\log F$ using the convolutional model \texttt{cflux} in \textsc{XSPEC}.\\
Moreover, in two GRBs (GRB090902B and GRB190114C) the prompt spectrum exhibits in NaI data a power law component \citep{Fermi2009,Ajello2020,Ursi2020} in addition to the main emission one, also at low energies. Therefore, we add a power law in both the Band and Synchrotron models, \texttt{powerlaw} in \textsc{XSPEC} notation. In these two cases, we compute the flux of the main component only (Band or Synchrotron), since the physical nature of this additional power law was already discussed in the literature and it is outside the scope of this work.\\  
For the Synchrotron case, we used the same table model presented in \cite{Oganesyan2019}. Synchrotron spectra are produced by a population of non-thermal electrons injected as a power law ${\rm d}N_e/{\rm d}\gamma \propto \gamma^{-p}$ with a characteristic minimum Lorentz factor $\gamma_m$. The final spectrum is obtained by convolving the single electron spectrum with the electron population distribution after that particles cooled through synchrotron losses down to the cooling Lorentz factor $\gamma_c$. Table spectra are produced by varying $\log\left(\gamma_m/\gamma_c\right)$ between -1 and 2, allowing for both fast ($\gamma_m > \gamma_c$) and slow ($\gamma_m < \gamma_c$) cooling regimes, and by varying $p$ between 2 and 5 and the normalization $N_{Sync}$ between 0 and $10^{20}$ (in cgs units). However, this table model assumes a constant cooling frequency $\nu_c = 1$ keV. To let this parameter vary, we multiply the table model with the \textsc{XSPEX} convolutional model \texttt{zmshift}, which shifts the spectrum along the energy axes. The resulting photon spectrum depends on four parameters: the ratio between the Lorentz factors $\gamma_m/\gamma_c$, the power law slope $p$ of the injected electron distribution, the spectral shift parameter $z_{shift}$ (univocally linked with the observed cooling frequency $\nu_c$), and the normalisation $N_{Sync}$.\\

\begin{table*}
	\begin{center}
		\begin{tabular}{c | c | c} 
			\hline\hline
			Model & Free Parameter & Prior \\
			\hline 
			& $\rm \log F\ [cgs] $ & $(-10,\ -3)$  \\%[1.5ex]
			& $\rm Calibration\ constant $ & $(0.7,\ 1.3)$  \\%[1.5ex]
                \hline 
			& $\rm \alpha$ & $(-6,\ 2)$ \\%[1.5ex]
			Band & $\rm \beta $ & $(-5,\ -1.9)$  \\%[1.5ex]
			& $\rm E_{ch}\ [keV]$ & $(8,\ 40000)$ \\%[1.5ex]
			\hline 
   			& $\rm z_{shift} $ & $(-0.999,\ 10)$ \\%[1.5ex]
			Synchrotron & $\rm p $ & $(2,\ 5)$  \\%[1.5ex]
			& $\rm \log (\gamma_m/\gamma_c) $ & $(-1,\ 2)$ \\%[1.5ex]			
                \hline 
   			& $\rm s $ & $(-5,\ 5)$ \\%[1.5ex]
			Power law &  &   \\%[1.5ex]
			& $\rm N_{PL}  $ & $(10^{-10},\ 10^{10})$ \\%[1.5ex]
   
			\hline\hline
		\end{tabular}
	\end{center}
	\caption{Free parameters and flat uninformative priors used for the spectral fit. Two main models, Band and Synchrotron, were fit to the dataset. For each of these models, we include cross-calibration constant and a measurement of the 10 keV - 10 MeV bolometric flux.}
	\label{tab:prior_spectral_fit}
\end{table*} 

\subsection{Parameter estimation, model comparison and goodness-of-fit}

For each free parameter of the two models, we define broad and  uninformative priors (Table \ref{tab:prior_spectral_fit}). The analysis returns posteriors of the parameters fitted to our dataset together with the Bayesian evidence.\\
Being also interested in physical parameters, we derived them from the inferred ones. In the case of the Band model, we evaluate the rest frame peak energy defined as $E^{Band}_{p,z} = (2+\alpha) \cdot E_{ch} \cdot (1+z)$, where $\alpha$ is the low-energy photon index and $E_{ch}$ is the Band characteristic energy.\\
In the case of the synchrotron model, we define the rest-frame synchrotron frequency associated to $\gamma_c$ as  $\nu_{c,z} = (1+z)/(1+z_{shift})$ and to $\gamma_m$ as $\nu_{m,z} = \nu_{c,z} \cdot (\gamma_m/\gamma_c)^2$ and the rest frame peak energy as $E^{Sync}_{p,z} = \max(\nu_{m,z}, \nu_{c,z})$. Here, $z_{shift}$ is the spectral shift parameter. \\
In both cases, we infer the logarithm of the 10 keV - 10 MeV rest frame flux $\log F$, from which we compute the flux $F = 10^{\log F}$ and the total isotropic-equivalent luminosity $L_{iso} = 4\pi D_L^2(z) \cdot F$, where $D_L(z)$ is the luminosity distance from the burst.\\
For each relevant parameter, we define the best fit value as the median on the posterior distribution, and lower/upper errors are derived from the 16th- and 84th-percentile of the posterior distribution, respectively.\\
To ensure the good convergence of the fit, we modified the prior of the cooling frequency for GRB170607A between $\nu_c \in (8,\ 500)$ keV, i.e. $z_{shift} \in (-0.998,\ -0.8)$.\\
From the Bayesian fit of both models to each GRB dataset, we get the Bayesian evidence $\log Z$. This provides a tool to compare which model describes better the data. In fact, the model that has a substantial larger evidence is the best-fit model. To assess this, we define the Bayes factor for these two models as $\log B = \log \left( Z_{Sync}/Z_{Band} \right)$, and we set the threshold to 0.5 according to Jeffrey's scale \citep{kass_raftery95}. If $|\log B| > 0.5$ we can identify a best-fit model, being the synchrotron one if $\log B > 0$ or the Band function if  $\log B < 0$. If $|\log B| < 0.5$, the Bayesian evidence of one model is not large enough to statistically exclude the other.\\
In addition, we aim to establish whether or not the synchrotron model provides a good fit to the data. To do so, we performed Posterior Predictive Checks (PPC) for each synchrotron fit to our sample.\\
By fitting a given model to a dataset, the best-fit model is the realization that maximizes the relative likelihood, therefore returning the smallest statistics. To asses whether that model realization is a good fit to the data, we simulate 1000 fake spectra out of the best-fit model, using the \textsc{XSPEC} command \texttt{fakeit}. We estimate, for each fake spectrum, the likelihood of the best-fit model out of the fake spectrum dataset.\\
By repeating this computation for all the 1000 fake spectra, we can define a (Gaussian-like) probability density function relative to the statistics distribution. We measure the goodness-of-fit through the comparison between the simulated statistics distribution and the measured likelihood value. Specifically, we compute the $p$-value of the real statistics on the simulated distribution, and we consider it a good fit if $p$-value $>0.05$. We define the $p$-value as the integral of the simulated statistics probability density function between the best-fit statistics value and $+\infty$ or $-\infty$ depending if the measured statistics is greater or lower the maximum of the distribution, respectively.\\

\subsection{Linear fit routine}

To investigate the Yonetoku relation, we fit the following power law relation to the data of our samples: 
\begin{equation}
 \dfrac{E_{p,z}}{100\ {\rm keV}} = K \cdot \left( \dfrac{L_{iso}}{10^{52}\ {\rm erg/s}}\right)^m  
 \label{eq:PLrelation}
\end{equation}
\\
Where $m$ and $K$ are the power law slope and the normalization, respectively. 
More practically, we use the logarithmic form of Eq. \ref{eq:PLrelation} in order to perform a linear fit:
\begin{equation}
 \log{E_{p,z}} = m\cdot \left(\log L_{iso} - 52 \right) + \log K - 2
 \label{eq:PLrelation_linear}
\end{equation}
\\
We normalise both \Ep\ and \Liso\ in order to perform the linear fit closer to the data barycenter, providing better $m$ and $K$ estimates. Values and $1\sigma$ errors for $\log E_{p,z}$ and $\log L_{iso}$ are derived directly from \Ep\ and \Liso\ parameter distributions, respectively.\\
We perform the fit using a Bayesian approach. 
We define a likelihood that assumes Gaussian parameter distributions for both $\log$\Ep\ and $\log$\Liso\ and takes into account errors on both parameters with the addition of an intrinsic Gaussian noise term $\sigma_{sc}$ \citep{D'Agostini2005}:

\begin{equation}
\begin{split}
    & - 2 \ln \mathcal{L}  \left( m, K, \sigma_{sc}\ |\ \{ x_i,\ \sigma_{x_i},\ y_i,\ \sigma_{y_i} \}\right) = \\
    & = \sum_{i = 0}^N \left[ \ln \left( \sigma^2_{sc} + \sigma^2_{y_i} + m^2 \sigma^2_{x_i}  \right) + \dfrac{(y_i - m(x_i - 52) - \log K+2)^2}{(\sigma^2_{sc} + \sigma^2_{y_i} + m^2 \sigma^2_{x_i} )} \right]
\end{split}
\end{equation}
where $N$ is the number of GRBs in a given sample, ($x_i$, $\sigma_{x_i}$) are $\log$\Liso\ mean and standard deviation relative to the $i$-th GRB and ($y_i$, $\sigma_{y_i}$) are $\log$\Ep\ mean and standard deviation relative to the $i$-th GRB, respectively. In this approach, together with $m$ and $K$ we estimate the intrinsic scatter $\sigma_{sc}$, which constitutes a new free parameter. \\
Given the shape of $\log$\Liso\ parameter distribution derived from the spectral fit, we note that the hypothesis of a Gaussian distribution constitutes a good approximation. For this reason, we define $\sigma_{x_i}$ as the quadrature sum of its upper and lower errors. On the other hand, the shapes of $\log$\Ep\ parameter distributions are asymmetric and deviate from a classic normal distribution. To account for this inconsistency, we still approximate $\log$\Ep\ as a Gaussian, but we consider $\sigma_{y_i}$ to be equal to the $\log$\Ep\ upper or lower error if the data point of the $i$-th GRB is above or below the best-fit line, respectively.\\
We adopt uniform priors for the three parameter, spanning the ranges 
$m \in (0,5)$, $K \in (0.01, 100)$ and $\sigma_{sc} \in (0,100)$. We sampled the posterior probability density with a Markov chain Monte Carlo approach using the \texttt{emcee} Python package \citep{emcee}, employing $N_{walk} = 32$ walkers. We initialized the walkers in a small 3-dimensional ball around a point in our parameter space $(m, K, \sigma_{sc}) = (0.5, 1.8, 10)$ and we performed $N_{iter}=5000$ iterations, for a total of $N_{iter} \times N_{walk} = 160000$ samples.\\
% Results are shown in Table \ref{tab:correlations_results}, testing different relations for different GRB samples.\\ 

\section{Results} \label{sec:results}

The main results of the spectral analysis fit to our \textit{single-bin} sample are shown in Tables \ref{tab:sync_sample} and \ref{tab:band_sample}.\\
From the Bayes factor comparison, we see that 21 GRBs are best fitted by Synchrotron model, 40 GRBs are best fitted by Band model while the two GRBs with a second component during the prompt emission peak prefer a Band model together with a power law. For 11 GRBs, the Bayes factor modulus $|\log B|$ is not large enough to prefer one model to the other.\\
However, the comparison between Band and the synchrotron models is not straightforward. Band is a phenomenological model, and it is able to accommodate a variety of spectra by providing measures of photon indices and peak energy. On the other hand, synchrotron model hinders many physical assumptions, providing precious insides on the nature of the emission, but still having less freedom in describing an observed spectrum.\\
Despite the Band function being the best-fit model for most of the GRBs in our sample, we notice that the synchrotron model provides a good fit to the data in the majority of the cases. For this reason we split the sample in two groups: the "Synchrotron" sample (with 58 GRBs out of 74, Table \ref{tab:sync_sample}) and the complementary "Band" sample (with 16 GRBs out of 74, Table \ref{tab:band_sample}). The latter includes all the GRBs that meet at least one of the following conditions: 1.) Synchrotron model provides a good fit of the data according to the PPC criterium; 2.) the best-fit model is Synchrotron; 3.) Bayes factor is not able to prefer one model to the other. For these GRBs we report parameter estimates obtained from the synchrotron fit.

\begin{longtable}[c]{cccccccc}
    \caption{List of bursts belonging to the \textit{single-bin} sample,  Synchrotron sub-sample. We show the results of the synchrotron fit for the burst that meet at least one of the following conditions: Synchrotron is the best-fit model, Synchrotron can not be excluded by the model comparison ($|\log B| < 0.5$), synchrotron model provides a good fit to the data. We report redshifts $z$ (see Sec. \ref{sec:sample_selection}), isotropic-equivalent luminosities \Liso, rest-frame peak energies \Ep\ (namely the minimum frequency $\nu_{m,z}$) and rest-frame cooling frequencies $\nu_{c,z}$ obtained from the Synchrotron fit. We report errors at 68 per cent confidence level. We further report the $p$-value returned from the PPC analysis, the Bayes factor $\log B$ obtained comparing Band and Synchrotron models, and the associated best-fit model: S = Synchrotron, B = Band and ND = Not Defined, i.e. when $|\log B| < 0.5$}
    \label{tab:sync_sample}\\

    \hline\hline \\
       GRB  & Redshift    & $L_{iso}$ &  $E_{p,z}$   & $\nu_{c,z}$ & $p$-value & $\log B$ & Best-fit model \\[0.5ex]
         &    & $[10^{51}\ \rm{erg/s}]$ &  [keV]   & [keV] &  &  &  \\[1ex]
         \hline \\
    \endfirsthead
    \multicolumn{8}{c}%
    {{\bfseries Table \thetable\ continued}} \\[0.5ex]
    \hline\hline \\
        GRB  & Redshift    & $L_{iso}$ &  $E_{p,z}$   & $\nu_{c,z}$ & $p$-value & $\log B$ & Best-fit model \\[0.5ex]
         &    & $[10^{51}\ \rm{erg/s}]$ &  [keV]   & [keV] &  &  &  \\[1ex]
         \hline \\
    \endhead
    
    GRB081222A & 2.77 & $127.74_{-8.9}^{+10.87}$    & $652_{-219}^{+385}$     & $340_{-86}^{+326}$   & 0.19     & -0.07         & ND            \\[0.5ex]
    GRB090323A & 3.57 & $228.3_{-18.94}^{+21.92}$   & $597_{-215}^{+430}$     & $288_{-73}^{+279}$   & 0.15     & 0.58          & S                 \\[0.5ex]
    GRB090328A & 0.74 & $13.09_{-0.96}^{+1.15}$     & $574_{-141}^{+284}$     & $381_{-72}^{+267}$   & 0.05     & -0.77         & B                 \\[0.5ex]
    GRB090424A & 0.54 & $19.18_{-0.32}^{+0.36}$     & $194_{-28}^{+41}$       & $157_{-11}^{+64}$    & 0.09     & -2.07         & B                 \\[0.5ex]
    GRB090618A & 0.54 & $20.5_{-0.92}^{+1.04}$      & $884_{-182}^{+189}$     & $161_{-16}^{+22}$    & 0.24     & 2.27          & S                 \\[0.5ex]
    GRB091003A & 0.9  & $44.97_{-2.04}^{+2.28}$     & $661_{-138}^{+274}$     & $482_{-68}^{+309}$   & 0.32     & -7.02         & B                 \\[0.5ex]
    GRB091127A & 0.49 & $5.63_{-0.2}^{+0.23}$       & $69_{-20}^{+45}$        & $40_{-6}^{+53}$      & 0.22     & 1.46          & S                 \\[0.5ex]
    GRB091208B & 1.06 & $20.44_{-1.59}^{+2.11}$     & $451_{-238}^{+249}$     & $74_{-14}^{+30}$     & 0.11     & 0.85          & S                 \\[0.5ex]
    GRB100414A & 1.37 & $96.46_{-5.56}^{+6.76}$     & $1051_{-314}^{+513}$    & $626_{-124}^{+427}$  & 0.35     & -0.72         & B                 \\[0.5ex]
    GRB100728A & 1.57 & $76.29_{-6.88}^{+8.34}$     & $1331_{-330}^{+649}$    & $859_{-191}^{+470}$  & 0.17     & -3.68         & B                 \\[0.5ex]
    GRB100816A & 0.8  & $8.54_{-0.48}^{+0.54}$      & $264_{-48}^{+98}$       & $204_{-28}^{+122}$   & 0.44     & -8.92         & B                 \\[0.5ex]
    GRB100906A & 1.73 & $53.64_{-4.19}^{+5.46}$     & $424_{-127}^{+263}$     & $246_{-58}^{+200}$   & 0.2      & -0.36         & ND                \\[0.5ex]
    GRB110213A & 1.46 & $21.96_{-1.54}^{+2.2}$      & $240_{-111}^{+120}$     & $54_{-16}^{+34}$     & 0.06     & 0.57          & S                 \\[0.5ex]
    GRB110731A & 2.83 & $181.07_{-8.97}^{+10.12}$   & $470_{-132}^{+232}$     & $273_{-57}^{+237}$   & 0.09     & 0.02          & NS            \\[0.5ex]
    GRB111228A & 0.71 & $4.29_{-0.3}^{+0.4}$        & $166_{-69}^{+60}$       & $22_{-6}^{+9}$       & 0.32     & 1.78          & S                 \\[0.5ex]
    GRB120119A & 1.73 & $77.09_{-6.05}^{+7.73}$     & $855_{-414}^{+649}$     & $240_{-55}^{+154}$   & 0.49     & 0.75          & S                 \\[0.5ex]
    GRB120624B & 2.2  & $302.1_{-20.9}^{+22.01}$    & $4047_{-2138}^{+2910}$  & $750_{-138}^{+279}$  & 0.23     & -0.03         & ND            \\[0.5ex]
    GRB120711A & 1.4  & $233.81_{-13.55}^{+12.71}$  & $2366_{-978}^{+3167}$   & $927_{-180}^{+337}$  & 0.25     & -1.8          & B                 \\[0.5ex]
    GRB120716A & 2.49 & $73.28_{-8.08}^{+10.11}$    & $544_{-234}^{+487}$     & $216_{-58}^{+191}$   & 0.34     & -0.41         & ND            \\[0.5ex]
    GRB121128A & 2.2  & $87.55_{-3.87}^{+4.43}$     & $339_{-69}^{+121}$      & $252_{-36}^{+145}$   & 0.12     & -2.58         & B                 \\[0.5ex]
    GRB130518A & 2.49 & $1082.21_{-27.5}^{+28.31}$  & $22123_{-1992}^{+2049}$ & $415_{-26}^{+26}$    & 0.33     & 23.26         & S                 \\[0.5ex]
    GRB131105A & 1.69 & $42.41_{-5.66}^{+6.39}$     & $4554_{-2262}^{+4878}$  & $63_{-15}^{+20}$     & 0.25     & -0.81         & B                 \\[0.5ex]
    GRB131108A & 2.4  & $287.82_{-12.67}^{+14.2}$   & $815_{-246}^{+583}$     & $471_{-73}^{+256}$   & 0.44     & -1.34         & B                 \\[0.5ex]
    GRB131231A & 0.64 & $35.71_{-1.13}^{+1.15}$     & $409_{-81}^{+151}$      & $295_{-35}^{+218}$   & 0.05     & 0.34          & ND            \\[0.5ex]
    GRB140213A & 1.21 & $27.14_{-0.77}^{+0.88}$     & $132_{-27}^{+50}$       & $94_{-13}^{+63}$     & 0.48     & 0.7           & S                 \\[0.5ex]
    GRB140506A & 0.89 & $9.63_{-0.95}^{+1.35}$      & $290_{-92}^{+194}$      & $159_{-42}^{+138}$   & 0.44     & -0.8          & B                 \\[0.5ex]
    GRB140512A & 0.72 & $11.0_{-1.11}^{+1.11}$      & $2696_{-1079}^{+1501}$  & $160_{-28}^{+41}$    & 0.44     & -1.41         & B                 \\[0.5ex]
    GRB140606B & 0.38 & $2.61_{-0.38}^{+0.41}$      & $5588_{-2818}^{+8437}$  & $32_{-7}^{+8}$       & 0.13     & -2.1          & B                 \\[0.5ex]
    GRB140808A & 3.29 & $147.53_{-10.0}^{+11.96}$   & $537_{-126}^{+274}$     & $369_{-77}^{+284}$   & 0.22     & -2.01         & B                 \\[0.5ex]
    GRB141028A & 2.33 & $258.37_{-14.54}^{+16.0}$   & $1306_{-458}^{+1102}$   & $649_{-142}^{+445}$  & 0.37     & -1.14         & B                 \\[0.5ex]
    GRB141220A & 1.32 & $34.3_{-2.01}^{+2.25}$      & $537_{-110}^{+232}$     & $393_{-63}^{+292}$   & 0.43     & -5.84         & B                 \\[0.5ex]
    GRB150403A & 2.06 & $466.87_{-25.65}^{+26.45}$  & $3498_{-1505}^{+1509}$  & $633_{-72}^{+151}$   & 0.14     & 2.44          & S                 \\[0.5ex]
    GRB150514A & 0.81 & $6.61_{-0.53}^{+0.58}$      & $103_{-28}^{+63}$       & $63_{-13}^{+48}$     & 0.03     & 0.1           & ND            \\[0.5ex]
    GRB150821A & 0.76 & $7.57_{-0.87}^{+1.03}$      & $1414_{-572}^{+783}$    & $86_{-15}^{+21}$     & 0.47     & 0.77          & S                 \\[0.5ex]
    GRB151027A & 0.81 & $6.38_{-0.58}^{+0.78}$      & $305_{-95}^{+212}$      & $176_{-48}^{+161}$   & 0.07     & -0.44         & ND            \\[0.5ex]
    GRB160509A & 1.17 & $204.26_{-4.04}^{+4.55}$    & $463_{-79}^{+128}$      & $357_{-24}^{+134}$   & 0.49     & -1.96         & B                 \\[0.5ex]
    GRB160625B & 1.41 & $1541.35_{-24.17}^{+24.18}$ & $4148_{-488}^{+439}$    & $946_{-44}^{+49}$    & 0.07     & 21.36         & S                 \\[0.5ex]
    GRB170214A & 2.53 & $418.75_{-26.99}^{+42.16}$  & $2412_{-1338}^{+7113}$  & $595_{-125}^{+252}$  & 0.2      & -0.17         & ND            \\[0.5ex]
    GRB170405A & 3.51 & $527.03_{-28.75}^{+30.83}$  & $1303_{-332}^{+864}$    & $831_{-147}^{+600}$  & 0.22     & -2.72         & B                 \\[0.5ex]
    GRB170607A & 0.56 & $2.76_{-0.25}^{+0.27}$      & $557_{-202}^{+219}$     & $19_{-7}^{+8}$       & 0.46     & 2.42          & S                 \\[0.5ex]
    GRB170705A & 2.01 & $107.16_{-5.16}^{+6.62}$    & $476_{-119}^{+235}$     & $311_{-60}^{+236}$   & 0.27     & 0.08          & ND            \\[0.5ex]
    GRB171010A & 0.33 & $5.55_{-0.15}^{+0.16}$      & $333_{-40}^{+36}$       & $42_{-3}^{+3}$       & 0.26     & 6.41          & S                 \\[0.5ex]
    GRB180703A & 0.67 & $12.38_{-1.44}^{+1.71}$     & $2680_{-1792}^{+2255}$  & $318_{-60}^{+111}$   & 0.42     & 0.53          & S                 \\[0.5ex]
    GRB180720B & 0.65 & $129.59_{-3.38}^{+3.67}$    & $3170_{-541}^{+619}$    & $268_{-12}^{+13}$    & 0.08     & 13.48         & S                 \\[0.5ex]
    GRB180728A & 0.12 & $0.71_{-0.01}^{+0.02}$      & $202_{-16}^{+12}$       & $7_{-1}^{+1}$        & 0.14     & 8.05          & S                 \\[0.5ex]
    GRB181020A & 2.94 & $537.56_{-34.57}^{+35.69}$  & $2113_{-371}^{+670}$    & $1638_{-229}^{+683}$ & 0.07     & -7.82         & B                 \\[0.5ex]
    GRB190324A & 1.17 & $19.11_{-0.91}^{+1.06}$     & $218_{-40}^{+83}$       & $168_{-25}^{+97}$    & 0.5      & -4.17         & B                 \\[0.5ex]
    GRB200524A & 1.26 & $30.08_{-3.08}^{+4.21}$     & $563_{-204}^{+441}$     & $275_{-63}^{+234}$   & 0.37     & -0.32         & ND            \\[0.5ex]
    GRB200613A & 1.23 & $39.91_{-1.2}^{+1.22}$      & $288_{-39}^{+64}$       & $238_{-18}^{+88}$    & 0.41     & -7.15         & B                 \\[0.5ex]
    GRB201216C & 1.1  & $104.02_{-4.86}^{+5.45}$    & $2287_{-420}^{+432}$    & $124_{-9}^{+9}$      & 0.23     & 4.97          & S                 \\[0.5ex]
    GRB210204A & 0.88 & $17.61_{-1.5}^{+1.45}$      & $951_{-240}^{+311}$     & $15_{-15}^{+6}$      & 0.36     & 0.89          & S                 \\[0.5ex]
    GRB210610B & 1.13 & $104.22_{-5.25}^{+6.0}$     & $1184_{-182}^{+320}$    & $960_{-106}^{+365}$  & 0.36     & -15.57        & B                 \\[0.5ex]
    GRB211023A & 0.39 & $2.25_{-0.12}^{+0.15}$      & $353_{-83}^{+82}$       & $16_{-3}^{+3}$       & 0.19     & 0.94          & S                 \\[0.5ex]
    GRB211211A & 0.08 & $2.87_{-0.04}^{+0.05}$      & $2408_{-104}^{+112}$    & $236_{-7}^{+6}$      & 0.0      & 66.52         & S                 \\[0.5ex]
    GRB220101A & 4.62 & $585.79_{-38.57}^{+43.3}$   & $1611_{-451}^{+886}$    & $1005_{-213}^{+811}$ & 0.25     & -2.51         & B                 \\[0.5ex]
    GRB220107A & 1.25 & $44.92_{-3.28}^{+4.11}$     & $898_{-225}^{+407}$     & $590_{-111}^{+371}$  & 0.14     & -2.34         & B                \\[0.5ex]
    GRB230204B & 2.14 & $78.49_{-9.14}^{+10.72}$    & $1270_{-474}^{+1156}$   & $568_{-172}^{+394}$  & 0.08     & -0.73         & B                \\[0.5ex]
    GRB230818A & 2.42 & $80.39_{-5.74}^{+7.1}$      & $575_{-147}^{+280}$     & $381_{-79}^{+269}$   & 0.33     & -1.48         & B                
\end{longtable}

\begin{longtable}[c]{cccccccc}
\caption{List of bursts belonging to the \textit{single-bin} sample, Band sub-sample. 
The GRBs here listed are the ones which do not meet the condition of Table \ref{tab:sync_sample}. Therefore, we report redshifts 
$z$ (see Sec. \ref{sec:sample_selection}), isotropic-equivalent luminosities \Liso, rest-frame peak energies \Ep\ and low energy photon indices $\alpha$ obtained from the Band fit. We report errors at 68 per cent confidence level. We further report the $p$-value returned from the PPC analysis, the Bayes factor $\log B$ obtained comparing Band and Synchrotron models, and the associated best-fit model: S = Synchrotron, B = Band and ND = Not Defined, i.e. when $|\log B| < 0.5$}
\label{tab:band_sample}\\
\hline\hline \\
   GRB  & Redshift    & $L_{iso}$ &  $E_{p,z}$   & $\alpha$ & $p$-value & $\log B$ & Best-fit model \\[0.5ex]
     &    & $[10^{51}\ \rm{erg/s}]$ &  [keV]   & &  &  &  \\[1ex]
     \hline \\
\endfirsthead
\multicolumn{8}{c}%
{{\bfseries Table \thetable\ continued}} \\[0.5ex]
\hline\hline \\
    GRB  & Redshift    & $L_{iso}$ &  $E_{p,z}$   & $\alpha$ & $p$-value & $\log B$ & Best-fit model \\[0.5ex]
     &    & $[10^{51}\ \rm{erg/s}]$ & [keV] &  &  &  &  \\[1ex]
     \hline \\
\endhead

GRB081221A & 2.26 & $125.09_{-5.62}^{+6.95}$    & $347_{-12}^{+15}$    & $-0.12_{-0.09}^{+0.08}$ & $2.8\times 10^{-5}$ & -21.01        & B             \\[0.5ex] 
GRB090102A & 1.55 & $58.69_{-4.3}^{+4.56}$      & $984_{-74}^{+73}$    & $-0.13_{-0.09}^{+0.11}$ & $2.2\times 10^{-2}$ & -15.69        & B             \\[0.5ex] 
GRB090510A & 0.9  & $248.08_{-15.72}^{+16.1}$   & $7842_{-546}^{+627}$ & $-0.69_{-0.04}^{+0.04}$ & $6.3\times 10^{-4}$ & -9.5          & B             \\[0.5ex] 
GRB090902B & 1.82 & $572.25_{-20.62}^{+22.39}$  & $2117_{-74}^{+83}$   & $0.18_{-0.08}^{+0.1}$   & $< 10^{-16}$ & -41.64        & B             \\[0.5ex] 
GRB090926A & 2.11 & $741.84_{-12.9}^{+14.39}$   & $778_{-25}^{+24}$    & $-0.4_{-0.03}^{+0.03}$  & $1.6\times 10^{-11}$ & -36.65        & B             \\[0.5ex] 
GRB140206A & 2.73 & $177.82_{-11.21}^{+12.83}$  & $450_{-27}^{+27}$    & $0.7_{-0.19}^{+0.22}$   & $1.6\times 10^{-11}$ & -33.19        & B             \\[0.5ex] 
GRB140508A & 1.03 & $128.89_{-4.0}^{+4.49}$     & $726_{-35}^{+35}$    & $-0.5_{-0.03}^{+0.03}$  & $2.3\times 10^{-2}$ & -14.73        & B             \\[0.5ex] 
GRB140801A & 1.32 & $30.65_{-1.17}^{+1.3}$      & $294_{-10}^{+10}$    & $-0.01_{-0.08}^{+0.08}$ & $1.0\times 10^{-4}$ & -29.85        & B             \\[0.5ex] 
GRB150314A & 1.76 & $394.64_{-10.69}^{+11.12}$  & $727_{-27}^{+27}$    & $-0.3_{-0.04}^{+0.04}$  & $1.1\times 10^{-16}$ & -42.04        & B             \\[0.5ex] 
GRB190114C & 0.42 & $68.9_{-1.48}^{+1.56}$      & $786_{-14}^{+14}$    & $-0.08_{-0.03}^{+0.03}$ & $< 10^{-16}$ & -294.36       & B             \\[0.5ex] 
GRB200826A & 0.75 & $23.76_{-1.81}^{+2.32}$     & $212_{-17}^{+17}$    & $-0.27_{-0.11}^{+0.13}$ & $3.1\times 10^{-3}$ & -7.67         & B             \\[0.5ex] 
GRB200829A & 1.25 & $724.37_{-12.25}^{+13.45}$  & $730_{-20}^{+24}$    & $-0.24_{-0.03}^{+0.04}$ & $< 10^{-16}$ & -88.57        & B             \\[0.5ex] 
GRB201020B & 0.8  & $25.65_{-1.62}^{+1.5}$      & $422_{-22}^{+29}$    & $-0.65_{-0.05}^{+0.04}$ & $4.0\times 10^{-2}$ & -2.33         & B             \\[0.5ex] 
GRB210619B & 1.94 & $2678.56_{-42.36}^{+40.39}$ & $1157_{-36}^{+38}$   & $-0.38_{-0.02}^{+0.02}$ & $< 10^{-16}$ & -88.54        & B             \\[0.5ex] 
GRB220527A & 0.86 & $48.13_{-1.18}^{+1.28}$     & $268_{-6}^{+6}$      & $-0.35_{-0.03}^{+0.03}$ & $2.5\times 10^{-14}$ & -52.47        & B             \\[0.5ex] 
GRB230812B & 0.36 & $50.44_{-0.41}^{+0.48}$     & $323_{-3}^{+3}$      & $-0.11_{-0.01}^{+0.01}$ & $< 10^{-16}$ & $< - 10^3$          & B             \\[0.5ex]
\end{longtable}

\clearpage

Interestingly, all the GRBs in this sample are in fast-cooling regime (\ratio\ $>1$), therefore in Table \ref{tab:sync_sample} we report \Liso, the peak energy \Ep\ $=\nu_{m,z}$ and the cooling frequency $\nu_{c,z}$, both in the rest-frame. On the other hand, the former sample includes all the GRBs where the synchrotron model does not provide a good fit to the data, and therefore we report parameter estimates obtained from the Band model fit.\\
In the Synchrotron sub-sample, all the spectra are well fitted by the Synchrotron model according to the PPC criterium except for three GRBs: GRB131231A, GRB150517A and GRB211211A. 
For the first two GRBs, $p-{\rm value} < 0.5$ because the best-fit statistics is particularly smaller than the average simulated statistics, i.e. it has a significantly larger likelihood. 
In the case of GRB211211A, the Bayes factor prefers the Synchrotron model despite the very large statistics compared to the simulated ones. In our synchrotron analysis, we get a comparable statistics with respect to the one reported by \citep{Gompertz2023} fitting a double 
smoothly broken power law function to the peak spectrum of this burst. \\
Because of model comparison, the results driven by analysing the \textit{single-bin} synchrotron sub-sample are valid under the assumption of synchrotron emission producing the prompt emission spectra.\\

\subsection{\Ep$-$\Liso\ relation}
\begin{table*}
    \caption{Results of the statistical analysis on the $E_{p,z} - L_{iso}$ and $\nu_{c,z} - L_{iso}$ relations relative to the \textit{single-bin} sample, to its Syncrotron sub-sample ("Sync"), to its Synchrotron sub-sample with intermediate-cooling spectra("Sync-Interm", i.e. $1< \nu_m/\nu_c < 3$) and to the \textit{multiple-bins} Synchrotron sub-sample. We report the Spearman's rank correlation coefficient $\rho$ and its associated chance probability $P_{chance}$, the slope $m$ and normalization $K$ of the power law fit and the intrinsic scatter $\sigma_{sc}$ of the data points around the best-fit line.}
    \begin{center}
         \begin{tabular}{c c c c c c c c}
            \hline\hline
                 Relation & Sample & $N_{GRB}$ & $\rho$  & $P_{chance}$ & $m$ & $K$ & $\sigma_{sc}$  \\
           \hline \\
                $E_{p,z} - L_{iso}$ & \textit{single-bin} & 74 & $0.49$ & $8.24\times 10^{-6}$ & $0.29 \pm 0.06$ & $4.44_{-0.66}^{+0.76}$ & $0.38$ \\\\  
                & \textit{single-bin} (Sync-Interm) & 35 & $0.74$ & $4.00\times 10^{-7}$ & $0.46_{-0.07}^{+0.08}$ & $2.33_{-0.34}^{+0.39}$ & $0.18$ \\
                \\ \hline \\
               $\nu_{c,z} - L_{iso}$ & \textit{single-bin} (Sync) & 58 & $0.81$ & $1.91\times 10^{-14}$ & $0.53 \pm 0.06$ & $0.97_{-0.13}^{+0.14}$ & $0.32$  \\ \\
                & \textit{multiple-bins} (Sync) & 77 & $0.83$ & $1.83\times 10^{-20}$ & $0.41 \pm 0.03$ & $1.12 \pm 0.09$ & $0.31$  \\ \\
               \hline
            \end{tabular}
    \end{center}   

    \label{tab:correlations_results}
\end{table*}
\begin{figure*}[ht!]
\centering 
	\includegraphics[width=0.49\textwidth]{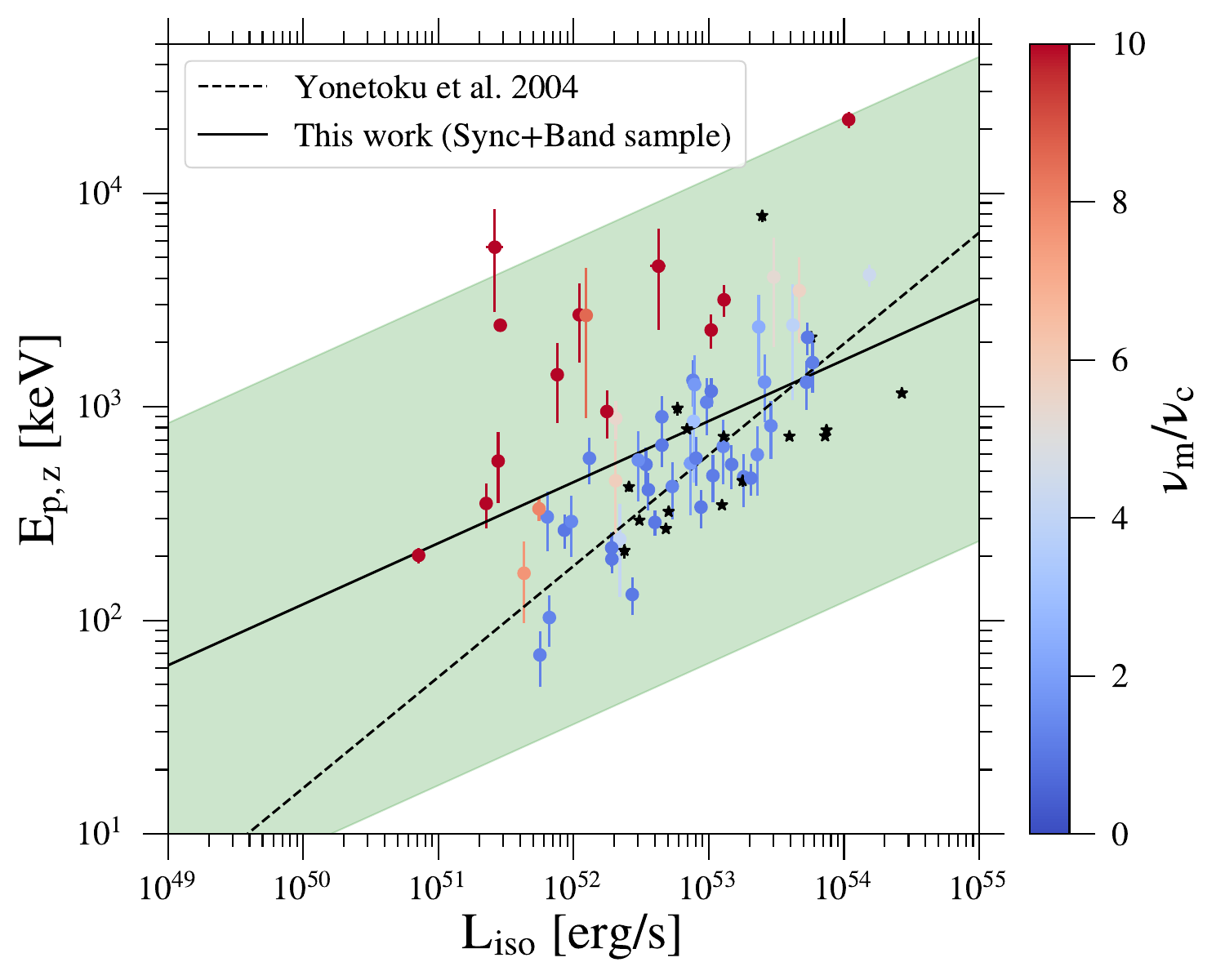}  \includegraphics[width=0.49\textwidth]{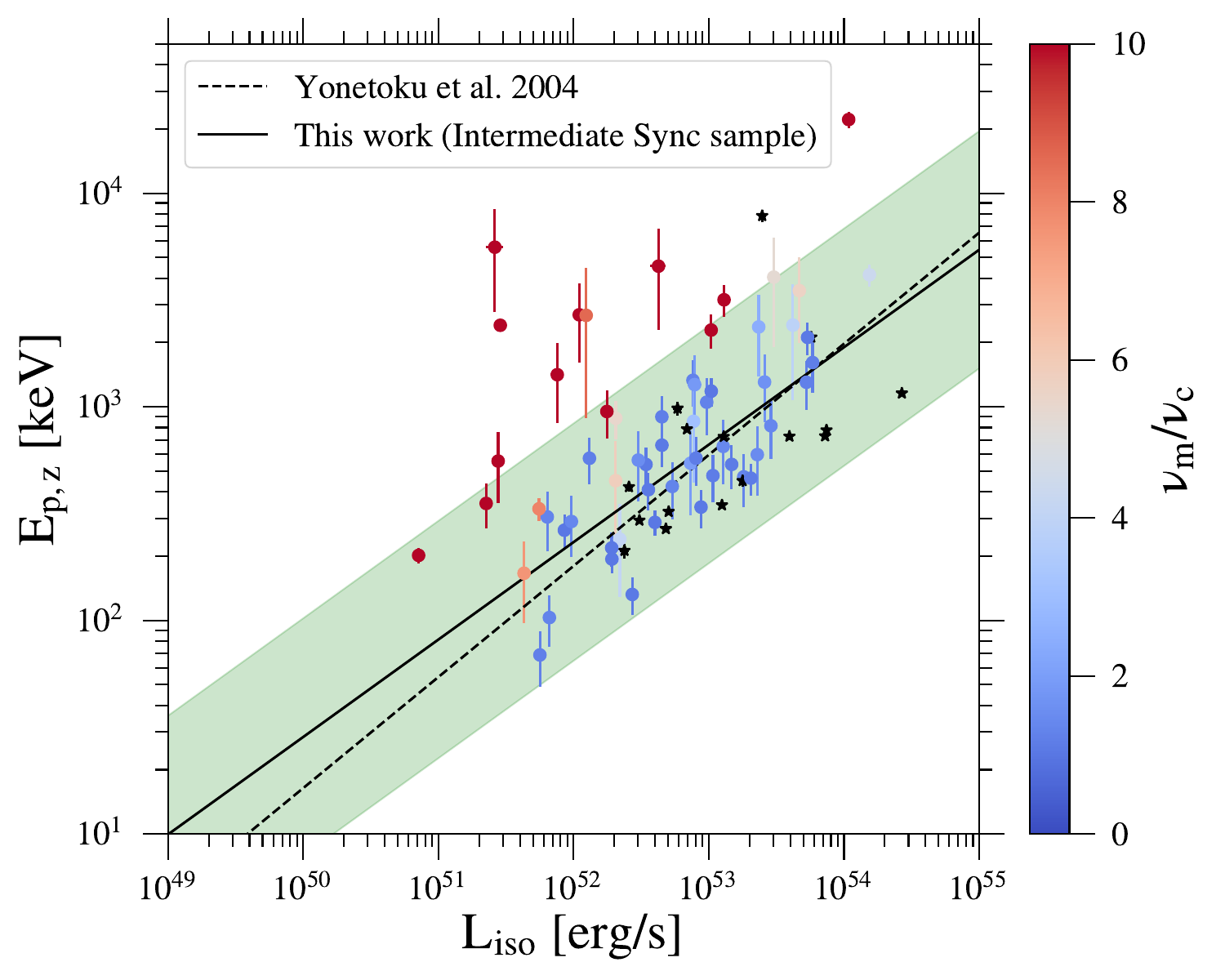}  
    \caption{Yonetoku relation obtained from the \textit{single-bin} sample. Black stars represent GRBs from the Band sub-sample, while colored dots represent GRBs from the Synchrotron sub-sample. The color scale is associated to the frequency ratio $\nu_m/\nu_c$ derived from the Synchrotron fit. We fit a power law to the whole \textit{single-bin} sample (left-hand panel) and to the GRBs in the Synchrotron sub-sample in an intermediate cooling regime (right-hand panel). The black straight line represents the best-fit line from the linear fit, and the green-shaded area the relative $3\sigma$ scatter region. For comparison, the black-dashed line represents the best-fit line from \citealt{Yonetoku2004}.}
	\label{fig:Yonetoku}
\end{figure*}

We first test the \Ep$-$\Liso\ relation obtained from the analysis of the whole sample. We computed the Spearman’s rank correlation coefficient $\rho$ and its associated chance probability $P_{chance}$.\\
We report the results in Table \ref{tab:correlations_results}, together with the ones of the linear fit. In Fig. \ref{fig:Yonetoku} (left panel) we show that the data are scattered around the best-fit line, suggesting the absence of any correlation between \Liso\ and \Ep\ ($\rho \sim 0.5$). In addition, the slope we obtain deviates from the one expected in the Yonetoku relation.\\
Nonetheless, we observe that GRBs with different frequency ratios \ratio\ populate differently the \Ep$-$\Liso\ plane. In fact, GRBs with a high ratio (i.e. in a fast-cooling regime, \ratio\ $\gtrsim 10$) tend to deviate from the Yonetoku relation, whereas GRBs with a lower ratio (i.e. in an \textit{intermediate}-cooling regime, \ratio\ $\lesssim 3$) tend to cluster along the Yonetoku relation.\\
By restricting the synchrotron sample to the GRBs in the intermediate-cooling regime ($1<$ \ratio\ $< 3$) and fitting the relation, we find that the data points are tightly correlated, and the linear fit returns values similar to the ones obtained in \citealt{Yonetoku2004} (Fig. \ref{fig:Yonetoku}, right panel). The results of this linear fit are also shown in Table \ref{tab:correlations_results}.\\

 \subsection{A physical relation: $\nu_{c,z} - L_{iso}$}

The \Ep$-$\Liso\ analysis previously discussed shows that, when the physical synchrotron model is used, the Yonetoku relation does not hold for the whole sample, but for only a subset of GRBs that show a synchrotron spectrum where $1<$ \ratio\ $<3$ (Fig. \ref{fig:Yonetoku}, left panel). This constitutes an important connection between the Yonetoku relation and a physical parameter, namely the ratio between the rest-frame cooling frequency $\nu_{c,z}$ and minimum frequency $\nu_{m,z}$. In particular, the GRBs exhibiting a relation between their \Ep\ and \Liso\ are the ones whose spectral break energy is particularly close to their peak energy. \\
Therefore, we test the possibility to have a correlation between the break energy, namely the cooling frequency $\nu_{c,z}$, and the isotropic-equivalent luminosity \Liso. We employ the same methodology used to investigate the \Ep$-$\Liso\ relation, but applied to the Synchrotron sub-sample of the \textit{single-bin} sample (Table \ref{tab:sync_sample}).
\begin{figure*}[ht!]
    \centering
    \includegraphics[width=0.65\linewidth]{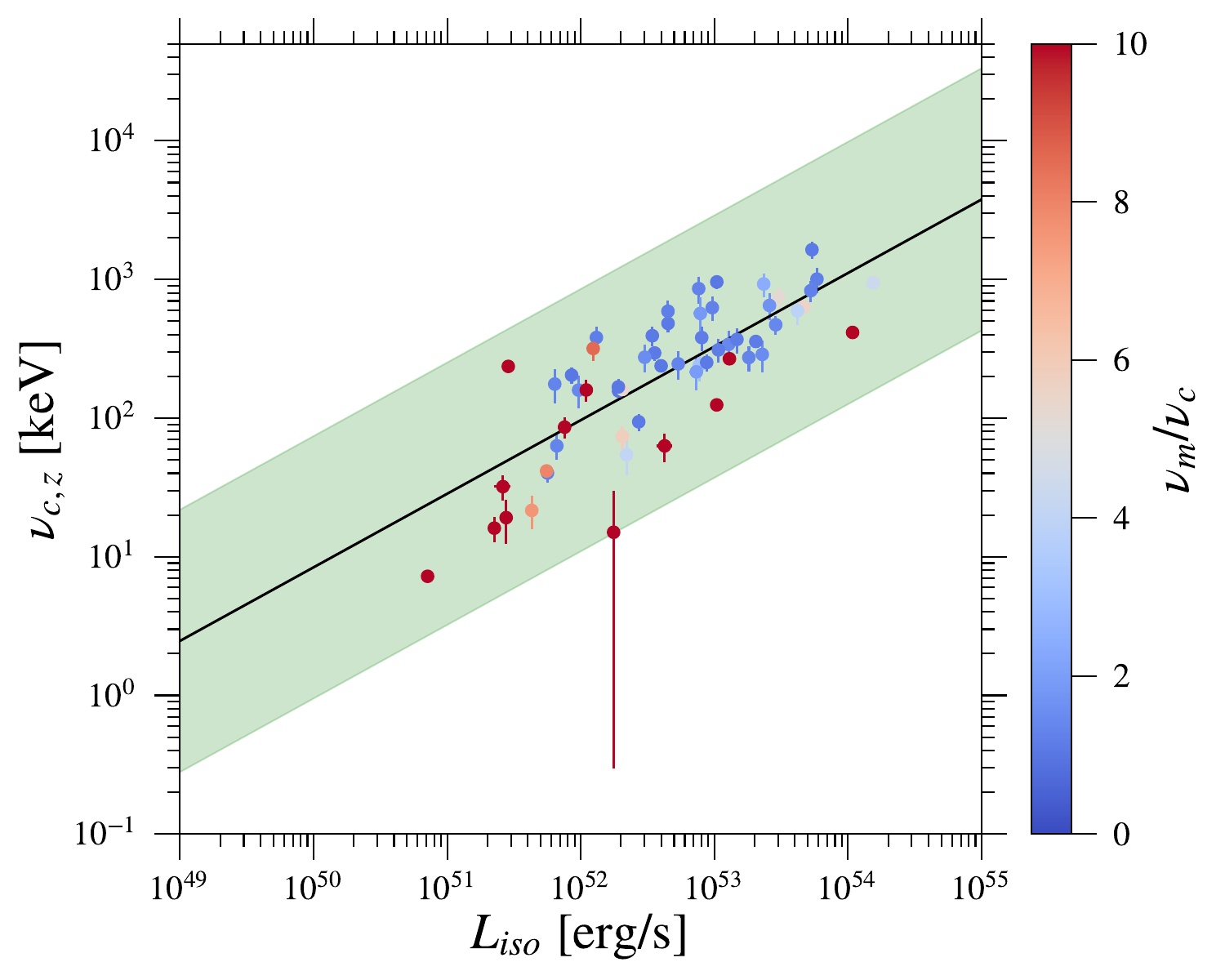}
    \caption{$\nu_{c,z} - L_{iso}$ relation obtained from the Synchrotron  \textit{single-bin} sample. The color scale is associated to the frequency ratio \ratio\ derived from the
    Synchrotron fit. The black straight line represents the best-fit line from the linear fit, and
    the green-shaded area the relative $3\sigma$ scatter region.}
    \label{fig:new_relation_allSample}
\end{figure*}
We find that all the GRBs in the Synchrotron sub-sample, regardless from their \ratio, show a tight $\nu_{c,z} - L_{iso}$ correlation (Fig. \ref{fig:new_relation_allSample}). Spearman's rank correlation test returns a coefficient $\rho = 0.81$ with $P_{chance} = 1.91 \times 10^{-14}$. The slope of the power law is consistent with both the one expected from the Yonetoku relation and the one from the \Ep$-$\Liso\ relation of GRBs in the Synchrotron sub-sample. The results are reported in Table \ref{tab:correlations_results}.\\

 \subsection{Multiple-bins analysis}

\begin{figure*}[ht!]
    \centering
    \includegraphics[width=0.65\linewidth]{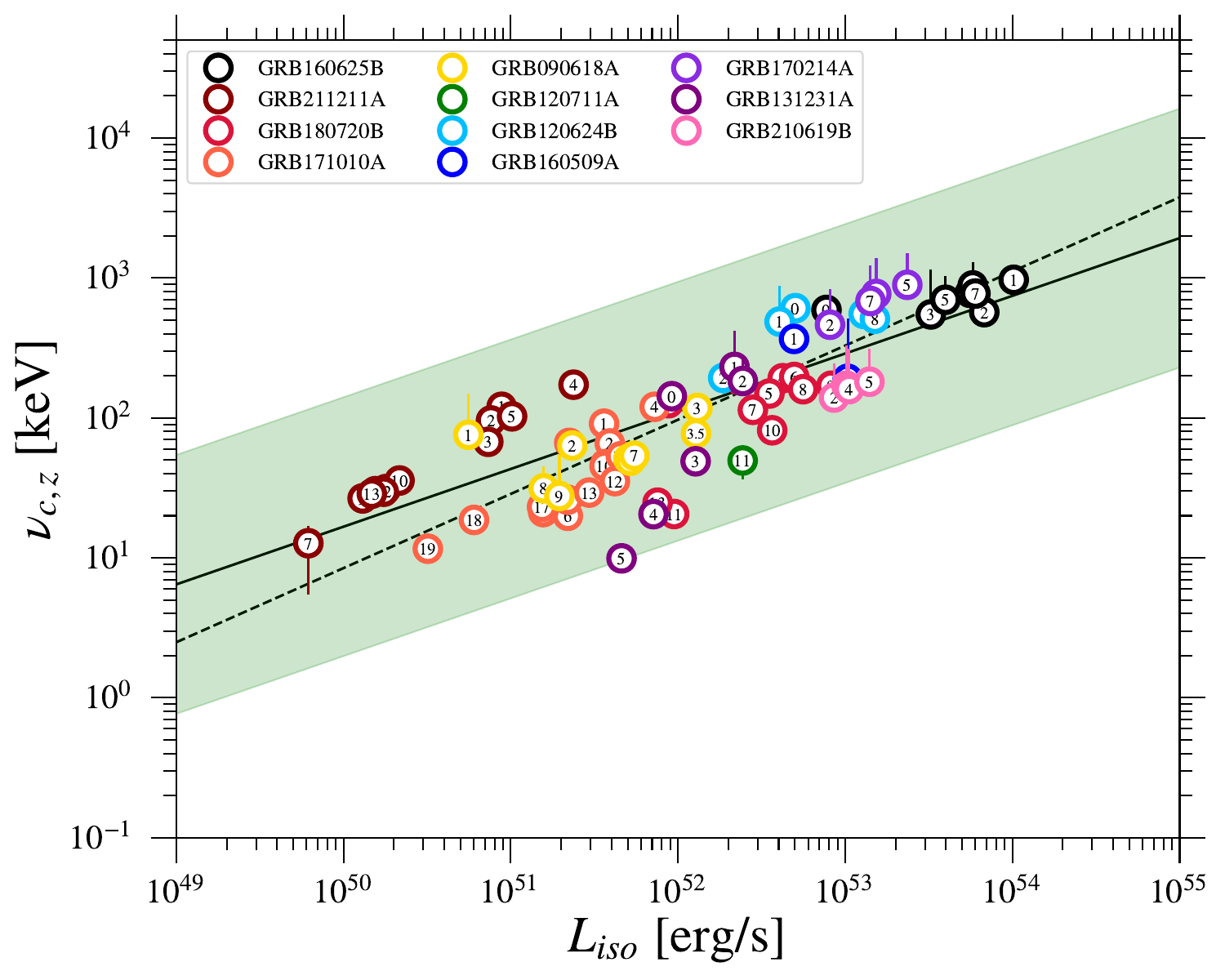}
    \caption{$\nu_{c,z} - L_{iso}$ relation obtained from the Synchrotron GBRs in the \textit{multi-bin} sample. We show multiple time-bins from different GRBs with circles of the same colors. The number inside the circles represents the bin number that data point is associated to. The black straight line represents the best-fit line from the linear fit while the dashed-line the best-fit line obtained from the $\nu_{c,z} - L_{iso}$ relation in the \textit{single-bin} sample. The green-shaded area shows the $3\sigma$ scatter region of the relation.}
    \label{fig:newRelation_multibinSample}
\end{figure*}

Out of the 74 GRBs in our total sample, we select 15 GRBs with the highest fluence (as reported by \gbm\ catalog) in order to perform time-resolved spectral analyses across their whole duration. Among these 15 GRBs we find GRB090902B and GRB190114C, namely the two GRBs exhibiting a second power law spectral component in addition to the main one. We remove these two GRBs from the sample, and the remaining 13 GRBs constitute the new \textit{multiple-bins} sample. The time-resolved spectral analysis relative to this sample includes both Band and synchrotron models, similarly to the procedure described in Sec. \ref{sec:methods}.\\
We visually selected the time-bins from their GBM light curve in order to match as much as possible a single GRB pulse. The time-bin width is chosen in order to collect enough signal for the spectral analysis. In case a GRB shows a particularly bright pulse, we select multiple time-bins associate to a single pulse. 
%Some of these GRBs have LLE data available. 
We include LLE spectral data in any time-bin where they are available. \\
The \textit{multiple-bins} sample covers a total of 125 spectra. The GRB list, their time-bins and the spectral fit results are reported in Table \ref{tab:timeresolved_sample}.\\
According to the Bayes factor, 65 spectra are best-fitted by Synchrotron model, 48 spectra are best-fitted by Band and for 12 spectra the Bayes factor can not determine which is the best-fit model. For the 48 spectra best-fitted by Band, we report the \Ep, \Liso\ and $\alpha$ estimates obtained from the Band model, whereas for the remaining 77 spectra we report \Ep, \Liso\ and $\nu_{c,z}$ obtained from the Synchrotron model.\\
Similarly to the analysis of the \textit{single-bin} sample, we observe that the fit with the Synchrotron model is not returning any \Ep $-$ \Liso\ relation. However, restricting the \textit{multi-bin} sample to its Synchrotron sub-sample (i.e. the 77 GRBs for which we report synchrotron results), we observe again a very tight correlation between $\nu_{c,z}$ and \Liso\ (Fig. \ref{fig:newRelation_multibinSample}). The Spearman's rank correlation coefficient is $\rho = 0.83$ with chance probability $P_{chance} = 1.83 \times 10^{-20}$ (Table \ref{tab:correlations_results}). 
We notice that the $\nu_{c,z} - L_{iso}$ relation derived in this way is described by a slightly flatter best-fit line, with slope $m = 0.41 \pm 0.03$ and normalization $K = 1.12 \pm 0.09$, whereas the intrinsic scatter is comparable ($\sigma_{sc} = 0.31$). It is worth to mention that, if GRB211211A time-resolved dataset is considered alone, is is consistent with a slope $m\sim 0.5$. If excluded from the \textit{multiple-bins} sample, the relative $\nu_{c,z} - L_{iso}$ relation has the same slope if the one obtained from the \textit{single-bin} one.\\

\clearpage
\clearpage
\begin{longtable}{cccccccc}
\caption{List of bursts belonging to the \textit{multiple-bins} sample. We report the bin number and the relative time interval with respect to the GBM trigger time. In column 4 we report the Bayes factor obtained comparing Synchrotron and Band models. In column 5 we report the best-fit model: S = Synchrotron, B = Band and ND = Not Defined, i.e. when $|\log B| < 0.5$. We report the isotropic-equivalent luminosities \Liso\ and rest-frame peak energies \Ep\ derived from the best-fit model. In column 8 we report the rest-frame cooling frequency $\nu_{c,z}$ or the low-energy photon index $\alpha$ depending if the best-fit model is Synchrotron or Band, respectively. If the best-fit model is not defined, we report $\nu_{c,z}$.}
\label{tab:timeresolved_sample}\\

\hline\hline \\
    GRB & Bin number & Bin interval  & $\log B$ & Best-fit model & \Liso & \Ep & $\nu_{c,z}$ [$\alpha$]\\[0.5ex]
         &  & [s from trigger] &   &  & [$10^{51}$ erg/s] & [keV] & [keV] \\[1ex]

    \hline \\

\endfirsthead
\multicolumn{8}{c}%
{{\bfseries Table \thetable\ continued}} \\[0.5ex]
\hline\hline \\
    GRB & Bin number & Bin interval  & $\log B$ & Best-fit model & \Liso & \Ep & $\nu_{c,z}$ [$\alpha$]\\[0.5ex]
         &  & [s from trigger] &   &  & [$10^{51}$ erg/s] & [keV] & [keV] \\[1ex]

     \hline \\
\endhead
GRB160625B & 0 & (185.0, 187.22) & 0.21 & ND & $77.26_{-3.69}^{+4.38}$ & $18075_{-2279}^{+3184}$ & $591_{-80}^{+104}$ \\[0.5ex]
 & 1 & (187.22, 189.44) & 37.44 & S & $1017.24_{-12.92}^{+15.28}$ & $6761_{-586}^{+518}$ & $970_{-35}^{+39}$ \\[0.5ex]
 & 2 & (189.44, 191.67) & 21.03 & S & $678.38_{-11.48}^{+10.77}$ & $2605_{-399}^{+366}$ & $570_{-25}^{+33}$ \\[0.5ex]
 & 3 & (191.67, 193.89) & 8.29 & S & $325.01_{-5.99}^{+6.14}$ & $941_{-268}^{+356}$ & $549_{-77}^{+599}$ \\[0.5ex]
 & 4 & (193.89, 196.11) & 0.83 & S & $557.32_{-8.24}^{+7.6}$ & $948_{-118}^{+182}$ & $783_{-38}^{+240}$ \\[0.5ex]
 & 5 & (196.11, 198.33) & 3.59 & S & $397.78_{-6.05}^{+6.62}$ & $895_{-147}^{+226}$ & $698_{-54}^{+333}$ \\[0.5ex]
 & 6 & (198.33, 200.55) & 2.28 & S & $578.69_{-8.65}^{+9.83}$ & $1131_{-184}^{+274}$ & $884_{-60}^{+412}$ \\[0.5ex]
 & 7 & (200.55, 202.78) & -0.17 & ND & $600.51_{-7.97}^{+8.05}$ & $935_{-123}^{+157}$ & $774_{-32}^{+213}$ \\[0.5ex]
GRB211211A & 0 & (0.0, 2.3) & -11.52 & B & $0.29_{-0.01}^{+0.02}$ & $1519_{-158}^{+156}$ & $-1.14_{-0.01}^{+0.02}$ \\[0.5ex]
 & 1 & (2.3, 3.6) & 20.07 & S & $0.88_{-0.02}^{+0.02}$ & $1815_{-145}^{+135}$ & $120_{-5}^{+4}$ \\[0.5ex]
 & 2 & (3.6, 5.0) & 25.04 & S & $0.76_{-0.02}^{+0.02}$ & $1371_{-121}^{+108}$ & $96_{-3}^{+4}$ \\[0.5ex]
 & 3 & (5.0, 6.2) & 21.46 & S & $0.73_{-0.02}^{+0.02}$ & $2117_{-192}^{+215}$ & $67_{-3}^{+3}$ \\[0.5ex]
 & 4 & (6.2, 7.4) & 79.99 & S & $2.37_{-0.04}^{+0.04}$ & $2612_{-131}^{+113}$ & $173_{-5}^{+5}$ \\[0.5ex]
 & 5 & (7.4, 10.0) & 68.62 & S & $1.02_{-0.02}^{+0.02}$ & $2341_{-114}^{+130}$ & $103_{-3}^{+3}$ \\[0.5ex]
 & 6 & (10.0, 11.67) & 7.85 & S & $0.18_{-0.01}^{+0.01}$ & $484_{-132}^{+150}$ & $31_{-2}^{+3}$ \\[0.5ex]
 & 7 & (11.67, 13.33) & 2.25 & S & $0.06_{-0.01}^{+0.01}$ & $353_{-183}^{+205}$ & $13_{-7}^{+4}$ \\[0.5ex]
 & 9 & (16.0, 18.0) & 4.14 & S & $0.13_{-0.01}^{+0.01}$ & $457_{-125}^{+122}$ & $27_{-3}^{+3}$ \\[0.5ex]
 & 10 & (18.0, 20.0) & 11.17 & S & $0.22_{-0.01}^{+0.01}$ & $717_{-111}^{+116}$ & $36_{-2}^{+2}$ \\[0.5ex]
 & 11 & (20.0, 22.0) & 6.97 & S & $0.15_{-0.01}^{+0.01}$ & $463_{-97}^{+79}$ & $30_{-2}^{+3}$ \\[0.5ex]
 & 12 & (22.0, 24.0) & 9.82 & S & $0.17_{-0.01}^{+0.01}$ & $505_{-73}^{+67}$ & $30_{-2}^{+2}$ \\[0.5ex]
 & 13 & (24.0, 26.0) & 6.69 & S & $0.15_{-0.01}^{+0.01}$ & $381_{-84}^{+75}$ & $29_{-2}^{+2}$ \\[0.5ex]
GRB180720B & 0 & (0.0, 1.79) & -0.22 & ND & $8.78_{-0.59}^{+0.57}$ & $8921_{-3300}^{+4387}$ & $127_{-19}^{+22}$ \\[0.5ex]
 & 1 & (1.79, 3.57) & -8.05 & B & $23.43_{-1.7}^{+1.9}$ & $2765_{-348}^{+384}$ & $-1.03_{-0.02}^{+0.02}$ \\[0.5ex]
 & 2 & (3.57, 5.36) & -2.48 & B & $28.4_{-1.75}^{+2.01}$ & $1934_{-182}^{+205}$ & $-0.95_{-0.02}^{+0.02}$ \\[0.5ex]
 & 3 & (5.36, 7.14) & 3.68 & S & $23.28_{-1.03}^{+0.95}$ & $3322_{-491}^{+544}$ & $186_{-13}^{+15}$ \\[0.5ex]
 & 4 & (7.14, 8.93) & 8.82 & S & $42.3_{-1.3}^{+1.33}$ & $4747_{-556}^{+521}$ & $192_{-9}^{+11}$ \\[0.5ex]
 & 5 & (8.93, 10.71) & 7.03 & S & $35.21_{-1.32}^{+1.21}$ & $3110_{-388}^{+413}$ & $149_{-8}^{+8}$ \\[0.5ex]
 & 6 & (10.71, 12.5) & 8.33 & S & $49.55_{-1.45}^{+1.42}$ & $3624_{-433}^{+414}$ & $196_{-9}^{+10}$ \\[0.5ex]
 & 7 & (12.5, 14.29) & 7.49 & S & $28.04_{-1.35}^{+1.27}$ & $1985_{-380}^{+413}$ & $114_{-7}^{+8}$ \\[0.5ex]
 & 8 & (14.29, 16.07) & 18.71 & S & $56.04_{-1.63}^{+1.66}$ & $3169_{-430}^{+452}$ & $160_{-6}^{+8}$ \\[0.5ex]
 & 9 & (16.07, 17.86) & 14.33 & S & $82.66_{-1.99}^{+2.17}$ & $2753_{-361}^{+448}$ & $170_{-6}^{+6}$ \\[0.5ex]
 & 10 & (17.86, 19.64) & 2.16 & S & $36.59_{-1.23}^{+1.26}$ & $1623_{-175}^{+234}$ & $82_{-4}^{+4}$ \\[0.5ex]
 & 11 & (19.64, 21.43) & 2.14 & S & $9.49_{-0.61}^{+0.71}$ & $697_{-219}^{+261}$ & $21_{-3}^{+3}$ \\[0.5ex]
 & 12 & (21.43, 23.21) & -0.57 & B & $8.47_{-0.63}^{+0.69}$ & $585_{-79}^{+95}$ & $-1.26_{-0.04}^{+0.04}$ \\[0.5ex]
 & 13 & (23.21, 25.0) & 2.73 & S & $7.55_{-0.48}^{+0.58}$ & $494_{-144}^{+135}$ & $25_{-3}^{+3}$ \\[0.5ex]
 & 14 & (28.0, 31.0) & -3.37 & B & $21.91_{-0.77}^{+0.9}$ & $790_{-45}^{+51}$ & $-1.08_{-0.02}^{+0.02}$ \\[0.5ex]
 & 15 & (49.0, 51.0) & -1.29 & B & $10.92_{-0.76}^{+0.87}$ & $826_{-77}^{+105}$ & $-1.04_{-0.03}^{+0.03}$ \\[0.5ex]
GRB171010A & 0 & (8.0, 14.0) & 4.58 & S & $2.24_{-0.12}^{+0.14}$ & $1501_{-340}^{+360}$ & $66_{-4}^{+5}$ \\[0.5ex]
 & 1 & (14.0, 21.0) & 8.4 & S & $3.61_{-0.12}^{+0.1}$ & $604_{-124}^{+149}$ & $91_{-5}^{+6}$ \\[0.5ex]
 & 2 & (21.0, 27.0) & 4.97 & S & $3.91_{-0.08}^{+0.1}$ & $407_{-54}^{+47}$ & $66_{-3}^{+3}$ \\[0.5ex]
 & 3 & (27.0, 30.0) & -4.21 & B & $3.38_{-0.13}^{+0.14}$ & $308_{-13}^{+13}$ & $-1.08_{-0.02}^{+0.02}$ \\[0.5ex]
 & 4 & (30.0, 35.0) & 11.32 & S & $7.26_{-0.13}^{+0.14}$ & $519_{-70}^{+65}$ & $120_{-5}^{+6}$ \\[0.5ex]
 & 5 & (35.0, 40.0) & 10.36 & S & $3.67_{-0.06}^{+0.08}$ & $303_{-32}^{+25}$ & $47_{-2}^{+2}$ \\[0.5ex]
 & 6 & (40.0, 45.0) & 13.1 & S & $2.19_{-0.03}^{+0.03}$ & $237_{-13}^{+11}$ & $20_{-1}^{+1}$ \\[0.5ex]
 & 8 & (45.0, 55.0) & 21.53 & S & $2.19_{-0.03}^{+0.04}$ & $292_{-19}^{+16}$ & $26_{-1}^{+1}$ \\[0.5ex]
 & 9 & (55.0, 60.0) & -6.66 & B & $4.79_{-0.12}^{+0.13}$ & $320_{-9}^{+9}$ & $-1.05_{-0.01}^{+0.01}$ \\[0.5ex]
 & 10 & (60.0, 65.0) & 6.72 & S & $3.64_{-0.07}^{+0.08}$ & $418_{-37}^{+28}$ & $45_{-2}^{+2}$ \\[0.5ex]
 & 11 & (65.0, 67.0) & 3.14 & S & $4.55_{-0.11}^{+0.11}$ & $328_{-34}^{+29}$ & $53_{-3}^{+3}$ \\[0.5ex]
 & 12 & (67.0, 70.0) & 13.74 & S & $4.2_{-0.07}^{+0.09}$ & $297_{-27}^{+21}$ & $35_{-2}^{+2}$ \\[0.5ex]
 & 13 & (70.0, 75.0) & 16.81 & S & $2.94_{-0.05}^{+0.05}$ & $236_{-16}^{+14}$ & $29_{-1}^{+1}$ \\[0.5ex]
 & 16 & (75.0, 80.0) & 12.84 & S & $1.56_{-0.03}^{+0.05}$ & $160_{-21}^{+15}$ & $21_{-1}^{+1}$ \\[0.5ex]
 & 17 & (80.0, 85.0) & 8.39 & S & $1.54_{-0.08}^{+0.11}$ & $224_{-53}^{+54}$ & $23_{-2}^{+2}$ \\[0.5ex]
 & 18 & (85.0, 100.0) & 9.62 & S & $0.6_{-0.02}^{+0.03}$ & $151_{-26}^{+21}$ & $19_{-1}^{+1}$ \\[0.5ex]
 & 19 & (130.0, 140.0) & 2.32 & S & $0.32_{-0.01}^{+0.01}$ & $141_{-21}^{+17}$ & $12_{-2}^{+2}$ \\[0.5ex]
GRB090618A & 0 & (0.0, 20.0) & -3.58 & B & $1.87_{-0.11}^{+0.14}$ & $322_{-16}^{+19}$ & $-0.58_{-0.06}^{+0.05}$ \\[0.5ex]
 & 1 & (20.0, 40.0) & 0.0 & ND & $0.56_{-0.03}^{+0.04}$ & $117_{-30}^{+57}$ & $76_{-15}^{+73}$ \\[0.5ex]
 & 2 & (50.0, 60.0) & 0.41 & ND & $2.33_{-0.12}^{+0.14}$ & $370_{-113}^{+96}$ & $63_{-8}^{+14}$ \\[0.5ex]
 & 3 & (60.0, 66.0) & 9.14 & S & $13.07_{-0.36}^{+0.35}$ & $903_{-97}^{+97}$ & $118_{-6}^{+7}$ \\[0.5ex]
 & 3.5 & (66.0, 70.0) & 0.92 & S & $12.81_{-0.35}^{+0.34}$ & $616_{-76}^{+70}$ & $77_{-5}^{+5}$ \\[0.5ex]
 & 4 & (70.0, 78.0) & 1.5 & S & $5.04_{-0.1}^{+0.12}$ & $290_{-35}^{+32}$ & $51_{-3}^{+4}$ \\[0.5ex]
 & 5 & (78.0, 83.0) & 1.87 & S & $5.18_{-0.13}^{+0.16}$ & $254_{-37}^{+34}$ & $49_{-4}^{+5}$ \\[0.5ex]
 & 6 & (83.0, 86.0) & -2.25 & B & $5.92_{-0.12}^{+0.15}$ & $253_{-10}^{+9}$ & $-1.08_{-0.03}^{+0.03}$ \\[0.5ex]
 & 7 & (86.0, 90.0) & 0.37 & ND & $5.48_{-0.15}^{+0.2}$ & $266_{-49}^{+42}$ & $54_{-5}^{+6}$ \\[0.5ex]
 & 8 & (107.0, 112.0) & 1.83 & S & $1.57_{-0.06}^{+0.09}$ & $117_{-46}^{+34}$ & $31_{-6}^{+14}$ \\[0.5ex]
 & 9 & (112.0, 117.0) & 2.29 & S & $1.95_{-0.07}^{+0.08}$ & $80_{-30}^{+24}$ & $28_{-5}^{+27}$ \\[0.5ex]
GRB120711A & 0 & (60.0, 66.0) & -2.46 & B & $52.28_{-4.11}^{+5.04}$ & $3273_{-416}^{+551}$ & $-0.87_{-0.05}^{+0.05}$ \\[0.5ex]
 & 1 & (66.0, 69.0) & -6.8 & B & $125.32_{-6.26}^{+7.13}$ & $3205_{-248}^{+296}$ & $-0.92_{-0.03}^{+0.03}$ \\[0.5ex]
 & 2 & (69.0, 72.0) & -2.68 & B & $139.82_{-8.16}^{+7.22}$ & $3215_{-320}^{+291}$ & $-0.91_{-0.03}^{+0.03}$ \\[0.5ex]
 & 3 & (72.0, 77.0) & -6.77 & B & $59.67_{-3.02}^{+3.44}$ & $1865_{-148}^{+182}$ & $-1.0_{-0.03}^{+0.03}$ \\[0.5ex]
 & 4 & (77.0, 80.0) & -2.44 & B & $50.65_{-5.25}^{+5.54}$ & $2593_{-473}^{+595}$ & $-1.06_{-0.05}^{+0.05}$ \\[0.5ex]
 & 5 & (80.0, 85.0) & -3.9 & B & $43.33_{-3.2}^{+3.83}$ & $2230_{-298}^{+363}$ & $-1.03_{-0.04}^{+0.04}$ \\[0.5ex]
 & 6 & (85.0, 89.0) & -4.38 & B & $83.85_{-5.24}^{+5.48}$ & $2329_{-212}^{+263}$ & $-0.93_{-0.03}^{+0.03}$ \\[0.5ex]
 & 7 & (89.0, 92.0) & -2.4 & B & $102.32_{-6.36}^{+6.84}$ & $2335_{-241}^{+254}$ & $-0.92_{-0.03}^{+0.04}$ \\[0.5ex]
 & 8 & (92.0, 97.0) & -9.83 & B & $151.99_{-5.74}^{+6.02}$ & $3269_{-252}^{+248}$ & $-0.93_{-0.02}^{+0.02}$ \\[0.5ex]
 & 9 & (97.0, 100.0) & -8.64 & B & $142.52_{-7.17}^{+6.95}$ & $3857_{-344}^{+345}$ & $-0.94_{-0.02}^{+0.03}$ \\[0.5ex]
 & 10 & (100.0, 105.0) & -7.66 & B & $101.72_{-4.52}^{+4.85}$ & $2623_{-225}^{+229}$ & $-0.95_{-0.02}^{+0.03}$ \\[0.5ex]
 & 11 & (105.0, 110.0) & -0.19 & ND & $24.36_{-2.66}^{+2.69}$ & $5095_{-2319}^{+3842}$ & $50_{-13}^{+14}$ \\[0.5ex]
GRB120624B & 0 & (-270.0, -245.0) & 0.38 & ND & $50.36_{-2.19}^{+2.4}$ & $22169_{-7168}^{+7351}$ & $608_{-75}^{+91}$ \\[0.5ex]
 & 1 & (-245.0, -225.0) & 0.36 & ND & $40.52_{-2.61}^{+2.83}$ & $1231_{-476}^{+966}$ & $488_{-108}^{+388}$ \\[0.5ex]
 & 2 & (-225.0, -200.0) & 0.33 & ND & $18.64_{-2.45}^{+2.6}$ & $3551_{-2101}^{+4415}$ & $193_{-43}^{+72}$ \\[0.5ex]
 & 3 & (-180.0, -160.0) & -0.59 & B & $13.12_{-3.02}^{+3.74}$ & $2720_{-1091}^{+2292}$ & $-0.98_{-0.12}^{+0.18}$ \\[0.5ex]
 & 4 & (-130.0, -118.0) & -1.67 & B & $47.0_{-2.91}^{+3.54}$ & $1374_{-117}^{+141}$ & $-0.66_{-0.06}^{+0.06}$ \\[0.5ex]
 & 5 & (-105.0, -94.0) & -4.64 & B & $74.92_{-4.55}^{+4.82}$ & $1415_{-101}^{+99}$ & $-0.6_{-0.05}^{+0.05}$ \\[0.5ex]
 & 6 & (-94.0, -80.0) & -4.71 & B & $120.32_{-4.37}^{+4.82}$ & $2167_{-125}^{+135}$ & $-0.64_{-0.03}^{+0.03}$ \\[0.5ex]
 & 7 & (4.0, 9.0) & 1.73 & S & $129.01_{-7.09}^{+7.06}$ & $10436_{-4465}^{+4346}$ & $554_{-77}^{+89}$ \\[0.5ex]
 & 8 & (9.0, 14.0) & 0.9 & S & $150.42_{-9.23}^{+9.07}$ & $3171_{-1355}^{+1533}$ & $511_{-67}^{+129}$ \\[0.5ex]
 & 9 & (14.0, 17.0) & -2.19 & B & $78.37_{-6.11}^{+7.68}$ & $1614_{-230}^{+325}$ & $-1.1_{-0.05}^{+0.05}$ \\[0.5ex]
GRB160509A & 0 & (0.0, 4.0) & -1.0 & B & $9.43_{-0.98}^{+1.27}$ & $1061_{-200}^{+288}$ & $-1.14_{-0.06}^{+0.06}$ \\[0.5ex]
 & 1 & (8.0, 10.0) & 1.73 & S & $49.42_{-3.75}^{+3.44}$ & $2894_{-1046}^{+1196}$ & $367_{-49}^{+75}$ \\[0.5ex]
 & 2 & (10.0, 13.0) & -1.55 & B & $126.85_{-2.54}^{+2.82}$ & $881_{-43}^{+48}$ & $-0.65_{-0.02}^{+0.03}$ \\[0.5ex]
 & 3 & (13.0, 16.0) & -2.56 & B & $140.18_{-2.07}^{+1.93}$ & $658_{-19}^{+19}$ & $-0.68_{-0.02}^{+0.02}$ \\[0.5ex]
 & 4 & (16.0, 18.0) & -1.27 & B & $175.86_{-3.28}^{+3.63}$ & $744_{-30}^{+36}$ & $-0.75_{-0.02}^{+0.02}$ \\[0.5ex]
 & 5 & (18.0, 20.0) & 2.06 & S & $103.57_{-3.55}^{+4.11}$ & $718_{-338}^{+455}$ & $190_{-29}^{+326}$ \\[0.5ex]
 & 6 & (20.0, 25.0) & -2.74 & B & $28.92_{-0.68}^{+0.84}$ & $511_{-32}^{+39}$ & $-1.01_{-0.03}^{+0.03}$ \\[0.5ex]
GRB170214A & 0 & (8.0, 15.0) & -1.2 & B & $116.24_{-8.3}^{+7.6}$ & $2136_{-228}^{+266}$ & $-0.9_{-0.04}^{+0.05}$ \\[0.5ex]
 & 1 & (15.0, 23.0) & 0.72 & S & $153.32_{-5.69}^{+6.57}$ & $1348_{-378}^{+540}$ & $769_{-128}^{+614}$ \\[0.5ex]
 & 2 & (23.0, 33.0) & -0.08 & ND & $81.1_{-3.28}^{+3.65}$ & $858_{-260}^{+437}$ & $463_{-91}^{+373}$ \\[0.5ex]
 & 3 & (33.0, 40.0) & -0.68 & B & $148.89_{-5.73}^{+6.56}$ & $1637_{-94}^{+103}$ & $-0.82_{-0.03}^{+0.03}$ \\[0.5ex]
 & 4 & (40.0, 44.0) & -1.86 & B & $216.88_{-8.14}^{+9.85}$ & $1689_{-108}^{+121}$ & $-0.7_{-0.04}^{+0.04}$ \\[0.5ex]
 & 5 & (44.0, 49.0) & 0.02 & ND & $235.13_{-7.59}^{+8.54}$ & $1377_{-362}^{+670}$ & $893_{-127}^{+604}$ \\[0.5ex]
 & 6 & (49.0, 53.0) & -2.1 & B & $205.64_{-6.56}^{+7.83}$ & $1341_{-105}^{+117}$ & $-0.65_{-0.04}^{+0.05}$ \\[0.5ex]
 & 7 & (53.0, 60.0) & 0.95 & S & $141.65_{-5.1}^{+6.08}$ & $1309_{-374}^{+624}$ & $685_{-122}^{+540}$ \\[0.5ex]
 & 8 & (60.0, 64.0) & -2.28 & B & $299.23_{-8.67}^{+8.73}$ & $1392_{-92}^{+107}$ & $-0.62_{-0.05}^{+0.04}$ \\[0.5ex]
 & 9 & (64.0, 68.0) & -1.52 & B & $212.33_{-7.39}^{+8.33}$ & $1251_{-112}^{+123}$ & $-0.65_{-0.05}^{+0.05}$ \\[0.5ex]
 & 10 & (68.0, 80.0) & -0.73 & B & $114.83_{-4.71}^{+4.05}$ & $1102_{-65}^{+67}$ & $-0.77_{-0.03}^{+0.03}$ \\[0.5ex]
 & 12 & (120.0, 140.0) & -4.15 & B & $50.35_{-2.56}^{+3.34}$ & $1071_{-81}^{+93}$ & $-1.06_{-0.03}^{+0.04}$ \\[0.5ex]
GRB131231A & 0 & (15.0, 20.0) & 2.81 & S & $9.19_{-0.37}^{+0.41}$ & $542_{-214}^{+176}$ & $143_{-15}^{+37}$ \\[0.5ex]
 & 1 & (20.0, 23.0) & 6.36 & S & $21.77_{-0.52}^{+0.54}$ & $470_{-139}^{+136}$ & $231_{-27}^{+187}$ \\[0.5ex]
 & 2 & (23.0, 26.0) & 3.34 & S & $24.4_{-0.53}^{+0.53}$ & $241_{-44}^{+78}$ & $184_{-16}^{+93}$ \\[0.5ex]
 & 3 & (26.0, 30.0) & 1.89 & S & $12.75_{-0.32}^{+0.31}$ & $268_{-36}^{+38}$ & $49_{-3}^{+3}$ \\[0.5ex]
 & 4 & (30.0, 35.0) & 0.83 & S & $7.14_{-0.11}^{+0.15}$ & $189_{-21}^{+13}$ & $21_{-1}^{+2}$ \\[0.5ex]
 & 5 & (35.0, 40.0) & 2.96 & S & $4.6_{-0.11}^{+0.16}$ & $190_{-24}^{+17}$ & $10_{-1}^{+2}$ \\[0.5ex]
GRB210619B & 0 & (-2.0, 2.5) & -125.36 & B & $1095.61_{-13.21}^{+12.81}$ & $1041_{-21}^{+23}$ & $-0.42_{-0.01}^{+0.02}$ \\[0.5ex]
 & 1 & (2.5, 10.0) & -9.25 & B & $306.85_{-5.59}^{+6.09}$ & $579_{-15}^{+17}$ & $-0.63_{-0.02}^{+0.02}$ \\[0.5ex]
 & 2 & (10.0, 23.0) & 1.49 & S & $85.75_{-3.1}^{+3.07}$ & $243_{-74}^{+139}$ & $140_{-17}^{+105}$ \\[0.5ex]
 & 3 & (35.0, 41.0) & 1.41 & S & $101.77_{-4.98}^{+5.63}$ & $345_{-119}^{+266}$ & $170_{-28}^{+160}$ \\[0.5ex]
 & 4 & (45.0, 49.0) & 2.08 & S & $105.53_{-3.85}^{+4.61}$ & $390_{-153}^{+158}$ & $161_{-25}^{+152}$ \\[0.5ex]
 & 5 & (49.0, 55.0) & 0.68 & S & $139.72_{-5.24}^{+5.91}$ & $279_{-73}^{+159}$ & $181_{-20}^{+132}$ \\[0.5ex]
GRB230812B & 0 & (-1.0, 1.0) & $< - 300$ & B & $37.5_{-0.4}^{+0.44}$ & $432_{-6}^{+6}$ & $-0.1_{-0.02}^{+0.02}$ \\[0.5ex]
 & 1 & (1.0, 2.0) & -242.09 & B & $43.53_{-0.44}^{+0.46}$ & $278_{-4}^{+4}$ & $-0.33_{-0.02}^{+0.01}$ \\[0.5ex]
 & 2 & (2.0, 4.0) & -2.94 & B & $8.66_{-0.1}^{+0.11}$ & $145_{-2}^{+2}$ & $-0.77_{-0.02}^{+0.02}$ \\[0.5ex]
GRB200829A & 0 & (16.0, 18.0) & -16.8 & B & $122.52_{-4.25}^{+4.41}$ & $812_{-47}^{+57}$ & $-0.31_{-0.07}^{+0.07}$ \\[0.5ex]
 & 1 & (18.0, 20.0) & -153.17 & B & $491.53_{-7.49}^{+8.11}$ & $735_{-17}^{+21}$ & $-0.15_{-0.03}^{+0.03}$ \\[0.5ex]
 & 2 & (20.0, 22.0) & -129.74 & B & $507.29_{-6.58}^{+7.11}$ & $630_{-15}^{+13}$ & $-0.24_{-0.03}^{+0.03}$ \\[0.5ex]
 & 3 & (22.0, 25.0) & -6.66 & B & $74.73_{-2.18}^{+2.4}$ & $340_{-15}^{+17}$ & $-0.49_{-0.07}^{+0.06}$
\end{longtable}
\clearpage

\section{Discussion}\label{sec:discussion}

\begin{figure*}
\centering 
	\includegraphics[width=0.49\textwidth]{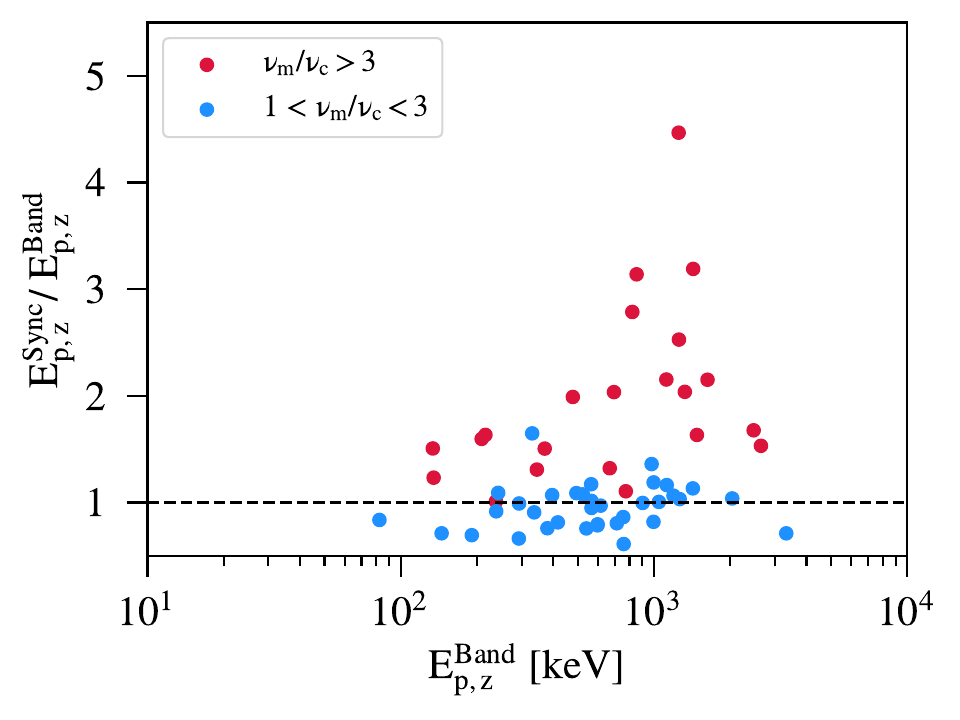}  \includegraphics[width=0.49\textwidth]{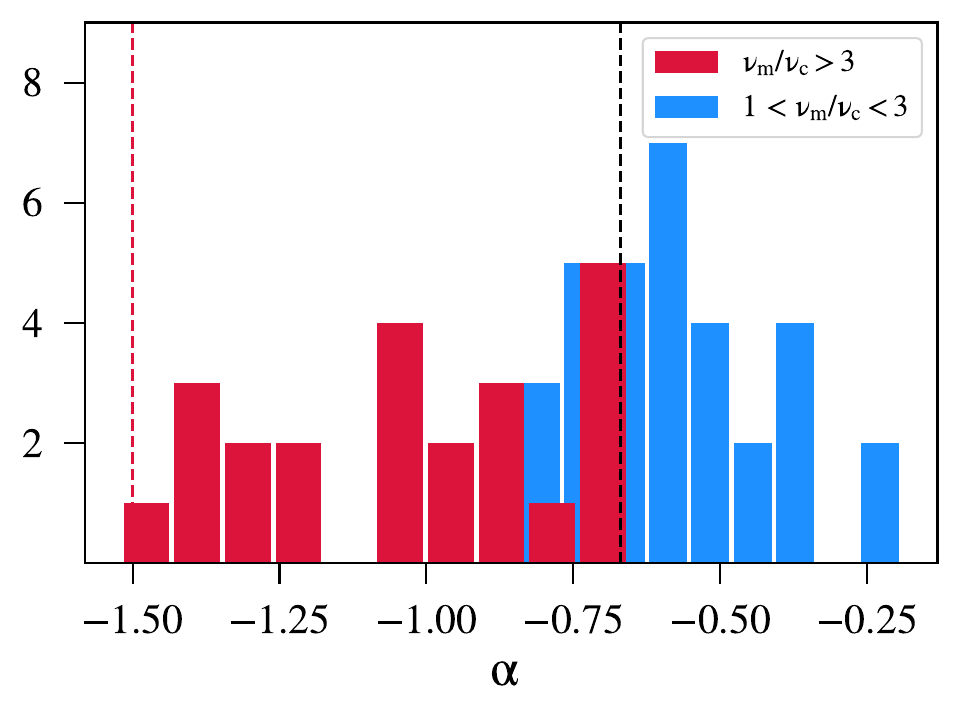}  
    \caption{Ratio between the rest-frame peak energy from Synchrotron model fits $E^{Sync}_{p,z}$ and Band model fits $E^{Band}_{p,z}$ as a function of $E^{Band}_{p,z}$ (left-hand panel) and low-energy break $\alpha$ histograms (right-hand panel), both relative to GRBs in the Synchrotron \textit{single-bin} sample. Parameters from GRBs associated with fast-cooling spectra (\ratio\ $>3$) are shown in red, while the ones associate to intermediate-cooling spectra ($1<$ \ratio\ $< 3$) in blue.}
	\label{fig:peak_alpha_comparisons}
\end{figure*}

\begin{figure*}
    \centering
    \includegraphics[width=0.7\linewidth]{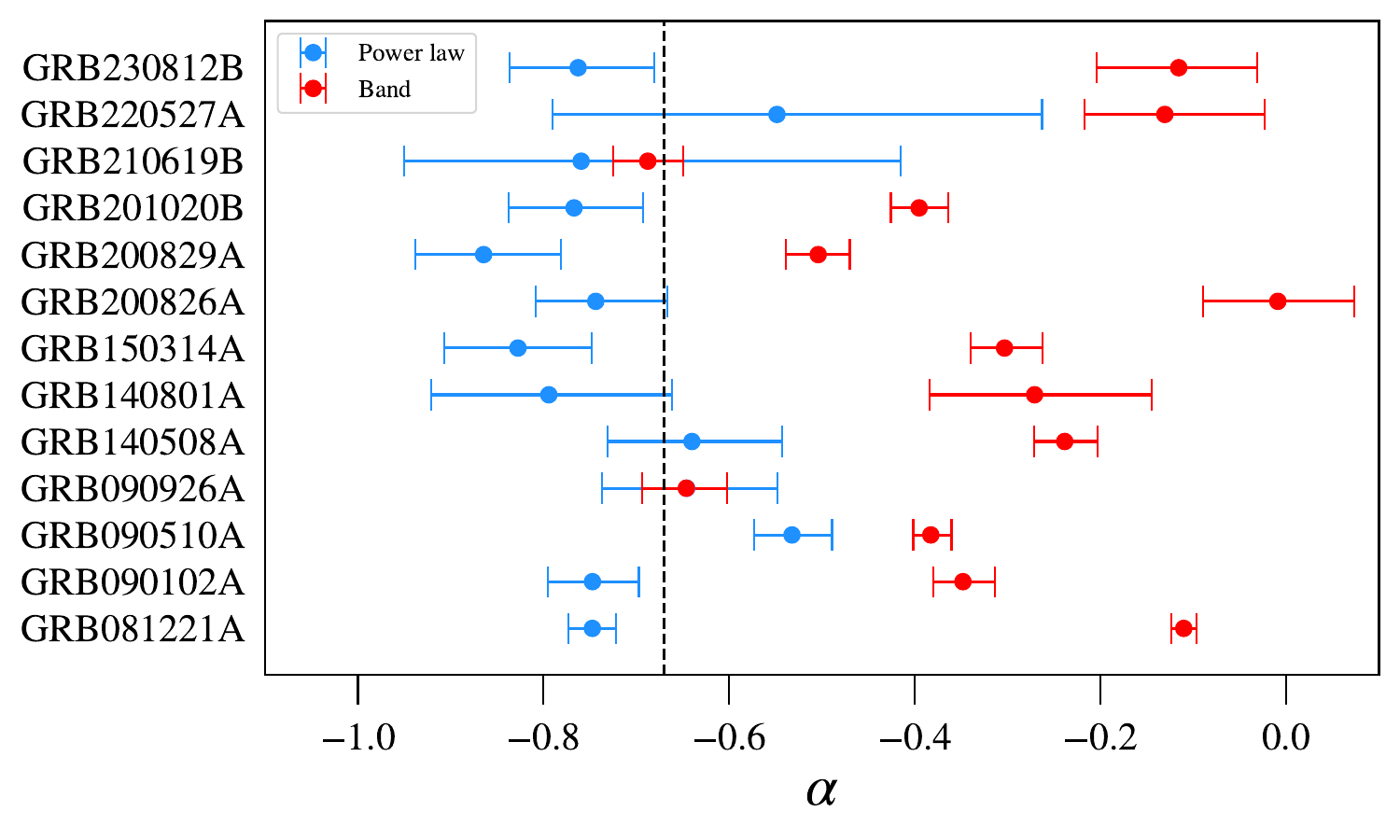}
    \caption{Low energy spectral index $\alpha$ obtained by fitting a Band model (in red) or a power law model considering data between 8-30 keV (in blue). The fit is performed to GRBs in the \textit{single-bin} Band sub-sample.
    The dashed vertical line corresponds to $\alpha = -0.67$, the value expected in synchrotron spectra for $E < E_{b}$. Errors are reported with 68$\%$ confidence level.}
    \label{fig:alphaPL}
\end{figure*}

We found a novel $\nu_{c,z}-L_{iso}$ relation by  employing a physical synchrotron model to describe the prompt emission of GRBs observed by \gbm. The fact that this new relation is found with high significance in the \textit{single-bin} and \textit{multiple-bins} samples, where both the emissions during the peak and across the whole burst duration are explored, strongly points towards its possible physical nature. In the following, we provide an interpretation of these results.\\
% The spectral peak energy \Ep\ may qualitatively represent a good proxy of GRB energetics. It constitutes a good indicator of the spectral hardness, it gives a measure of the bulk of the radiated energy and, considering GRBs with the same bolometric photon flux, its increase naturally leads to brighter bursts.\\
In the internal shocks framework, synchrotron spectra have a peak energy \Ep\ that depends on many other parameters, most relevantly the bulk Lorentz factor $\Gamma$ and the dynamical timescale $t_{var}$ \citep{Rees&Meszaros2005}. The presence of a \Ep$-$\Liso\ correlation would imply that $\Gamma$ and $t_{var}$ are the same for all the GRBs. This condition in not supported by any observation on large GRB samples. Therefore, this scenario is not in accordance with the Yonetoku relation, and our findings point towards this conclusion.\\
As a matter of fact, when time-resolved spectral analysis is performed, \Ep\ exhibits a strong spectral evolution, which can track the intensity of the pulse (e.g. \citealt{Golenetskii1983}) or show a hard-to-soft transition (e.g. \citealt{Lu2012}). The latter case is not consistent with the Yonetoku relation when multiple pulses are present. On the other hand, \cite{Ravasio2019} performed time resolved analysis of a sample of 10 bright long GRBs, fitting their spectra with a double smoothly broken power law (2SBPL) function, namely a Band function with two power laws before the peak. They find that the break energy, consistent with the synchrotron break in a marginally fast-cooling regime, does not vary much across the GRB duration. This finding suggests that the break energy may be more "standard" than the peak energy, or at least have minor dependencies on other parameters. However, the values of peak and break energies obtained through 2SBPL fits do not necessarily coincide with synchrotron breaks due to the shape of synchrotron spectra.\\
The question arises whether how the $\nu_{c,z}-L_{iso}$ relation can agree with almost two decades of observations pointing towards correlations with the peak energy \Ep.\\ 
A synchrotron SED in fast-cooling regime (\ratio\ $>1$) exhibits two typical energies: a peak around $\nu_m$ and a break around $\nu_c$. Before the peak, we expect a hard photon index $\alpha_1 = -2/3$ for $E < h\nu_c$, while for $h\nu_c < E < h\nu_m$ the emission softens with a photon index $\alpha_2 = -3/2$. The smaller the frequency ratio \ratio\ is, the closer the two breaks are. \\
Historically, most of the spectral fits that lead to the Yonetoku relation were carried out using phenomenological models which allowed for only one power law segment at low energies, with a single spectral break coinciding with the peak. In the hypothesis that GRBs prompt spectra are produced by synchrotron processes, the fit of a single-break model (such as the Band function) to a spectrum that intrinsically presents two breaks (such as a synchrotron one) can lead to two main observational biases: 

\begin{enumerate}
    \item[1.)] It appears that, when a single-break model like Band is fitted to an intrinsic double-break model, it returns $E_{b} < E^{Band}_{p} < E_{p}$, where $E^{Band}_{p}$ is the Band peak energy whereas $E_{b}$
    and $E_{p}$ are break and peak energies. Therefore, the fit of a two-breaks model to a suitable spectrum leads to systematically larger \Ep.  \\

    \item[2.)] The low-energy spectral index $\alpha$ that one obtains fitting a single-break model is a weighted average of the two spectral indices $\alpha_1$ and $\alpha_2$ before and after the break, respectively \citep{Toffano2021}. Therefore, the hardness of the spectrum depends on the proximity of $E_{b}$ to $E_{p}$. 
    
\end{enumerate}
By comparing results from the Band and Synchrotron models related to the \textit{single-bin} Synchrotron sample, we observe a similar pattern. In fact, when the synchrotron fit returns a fast-cooling spectrum, its peak energy $E^{Sync}_{p,z}$ is systematically larger than the value obtained by fitting the Band model $E^{Band}_{p,z}$. Conversely, when  $1 < $\ratio $< 3$ the two frequencies are very close and break and peak energies are hardly distinguishable, Band and synchrotron fits return similar peak energies, i.e. $E^{Sync}_{p,z}/E^{Band}_{p,z} \sim 1$ (Fig. \ref{fig:peak_alpha_comparisons}, left panel).
\\
In Fig. \ref{fig:peak_alpha_comparisons} (right panel) we show the distributions of the photon index $\alpha$ associated to the fast-cooling and intermediate-cooling spectra. While fast-cooling spectra span a large range of values pointing to softer spectra, the intermediate-cooling spectra exhibit harder photon indices close to $\alpha = -0.67$, which is the value expected for $E < E_{b,z} = h\nu_{c,z}$. \\
We can conclude that the effect of fitting a single-break model such as the Band function to a synchrotron spectrum leads to systematically lower values of \Ep. These are the premises by which the Yonetoku relation was vastly studied in literature. According to these findings, it is not surprising that when a physical synchrotron model is used in testing the Yonetoku relation, the latter holds only for GRBs where $1<$ \ratio\ $< 3$, namely when peak and break energies nearly coincide and therefore Band fits return correct \Ep\ estimates. This suggests that, at least for the GRBs that can be correctly described by an idealised synchrotron model, the underlying physical relation that holds is the $\nu_{c,z}-L_{iso}$ one. When spectra are analysed with the Band function, $E^{Band}_{p,z}$ is not far from $h\nu_c$, returning the notorious Yonetoku relation. However, when a second break is accounted for, the Yonetoku relation does not hold anymore, and the new $\nu_{c,z}-L_{iso}$ appears to be the underlying relation.
\begin{figure}
    \centering
    \includegraphics[width=1\linewidth]{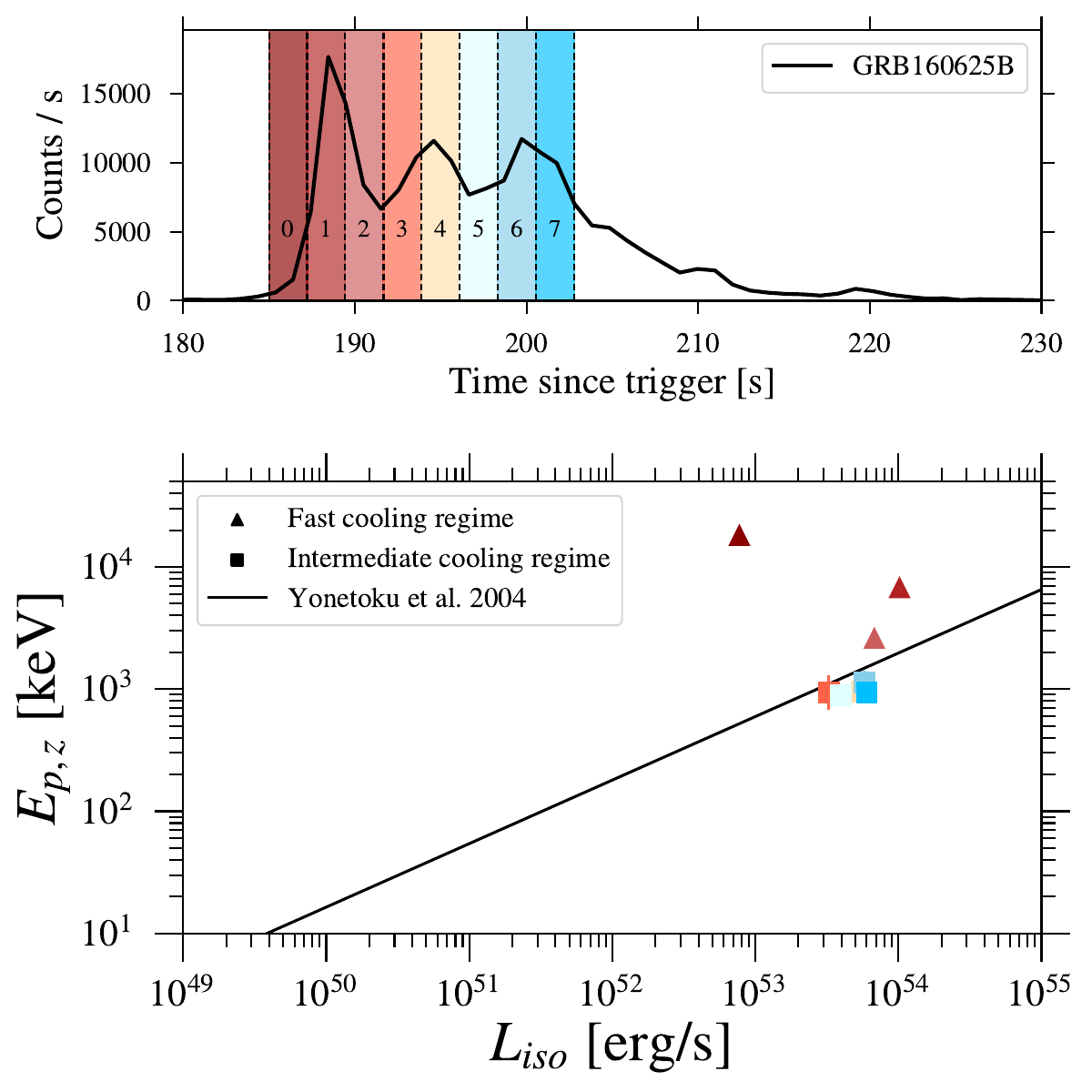}
    \caption{\gbm\ (8 - 900 keV) light curve and time-bins of the associated time-resolved analysis of GRB160625B (upper panel). In the lower panel, we show the relative \Ep$-$\Liso\ plane. The color-scale is associated to the time-bin in which the spectral analysis is performed. We show, for this GRB, the results of the Synchrotron fit, and we represent the bins with fast-cooling spectra as triangles, whereas the bins with intermediated-cooling spectra as squares. The black straight line represents the Yonetonu relation obtained from \cite{Yonetoku2004}.}
    \label{fig:160625_example}
\end{figure}\\
In Fig. \ref{fig:160625_example} we show an example of this phenomenon for GRB160625B, one of the brightest sources in the \textit{multiple-bins} sample. Its time-resolved analysis shows that, during the time-bins where the spectrum is in fast-cooling regime, the corresponding \Ep\ and \Liso\ are outliers of the Yonetoku relation. As soon as its spectrum shows an intermediate-cooling state, its \Ep\ approaches the relative $\nu_{c,z}$ and the corresponding \Ep\ and \Liso\ follow the Yonetoku relation.\\
Nonetheless, \cite{Guiriec2013, Guiriec2015} discuss an opposite scenario. They fit a two components model to time-resolved GRB prompt spectra. The presence of a second peaked component at lower energies resembles (in the photon flux representation) a total spectrum with two breaks. Despite finding systematically larger \Ep\ with respect to single-break model fits, they claim a tighter Yonetoku relation when a second component is accounted for in the fit. This behaviour was proven in a sample of 5 GRBs (e.g. \citealt{Guiriec2015b}). However, the extension of this to larger samples would be in contrast with the findings of \Ep $-$ \Liso\ relations in numerous samples when single-break models are used. In fact, if both the \Ep\ increase and the Yonetoku relation from single-break models are valid observational evidences, the fit of a second thermal component (or a double-break model) would systematically bring GRBs out of the relation. In this scenario, the only way to have a tighter relation would be for these GRBs exhibiting breaks to be all systematically beneath the best-fit line, i.e. to be scattered away from the best-fit relation with systematically lower \Ep\ for the same \Liso. Nonetheless, this hypothesis is hard to test in a small sample. In this work, testing a large sample of GRBs, we instead observed that GRBs in synchrotron sub-samples do not preferentially lay underneath the best-fit line, and this leads to the absence of the Yonetoku relation in both our samples.\\
It is important to stress that the results derived from the \textit{single-bin} sample are obtained under the assumption of synchrotron being the main radiative process producing prompt emission. This hypothesis is verified through model comparison only for 21/74 GRBs. On the other hand, as also discussed in \cite{Toffano2021}, the comparison between physical and phenomenological models is quite tricky, being the peculiar synchrotron features lost in low signal-to-noise spectra, often favouring a Band function. However, the conclusions that we draw remain valid in this working hypothesis, caveat providing a good synchrotron fit therefore reliable estimates of $\nu_m$ and $\nu_c$. Nevertheless, the results drawn from the \textit{multiple-bins} sample go beyond this assumption, selecting only Synchrotron spectra through model comparison.\\
In this regards, one important result of this work is the fact that the synchrotron model we used to fit the data is not able to well describe all the spectra. The interpretation can be twofold: or it appears that prompt dissipation sites and/or radiative processes are not the same for all the GRBs (therefore synchrotron emission can not explain all GRBs spectra), or that the synchrotron model we use is too simplistic to account for the GRB spectra. \\
From the \textit{single-bin} Band sub-sample (Table \ref{tab:band_sample}), we notice that the spectra that reject the synchrotron model show a relatively hard spectral index $\alpha$. The presence of these hard prompt emission spectra are in tension with a synchrotron interpretation, pointing to other emission mechanisms possibly produced in the optically thick region. However, we perform an additional fit of GRB spectra in the \textit{single-bin} Band sub-sample restricting the dataset between 8-30 keV and fitting a power law model. We exclude from this analysis GRB190114C and GRB090902B, since their low energy spectrum at few keV is dominated by a second power law component.\\
We observe that the spectral indices obtained through the power law fit $\alpha_{PL}$ is much smaller, i.e. the spectra are softer, than the ones obtained through Band fits $\alpha_{Band}$ (Fig. \ref{fig:alphaPL}). 
In particular, the $\alpha_{PL}$ values obtained are close to the value $\alpha = -0.67$ expected from synchrotron spectra at energies $E < E_{b}$. The reason of this mismatch can be due to the fact that the photon index $\alpha$ of the Band function is degenerate with the smoothness and position of the spectral peak. This finding suggests that $\alpha$ may be a biased indicator of GRB hardness \citep{Burgess2015}.\\
Still, the synchrotron model used in this work is not able to provide a good representation on the data for 13 GRBs. We propose that this fact is not due to the violation of the single electron spectrum, as often stated in literature, but rather to the shape and smoothness of the spectral profile around the peak. In these regards, there are many process not accounted for in our modelling, such as inverse Compton scattering in Klein-Nishina regime \citep{Derishev2009, Bosnjak2009, Daigne2011} provided by decaying magnetic fields \citep{Pe'er2006,Derishev2007, Zhao2014, Daigne2024}, anisotropic pitch angle distribution \citep{Medvedev2000,Sobacchi2021}  and so on. Accounting for all these process may lead to spectra more in accordance with the observations, without the need of introducing different radiative processes.\\
When instead also a simple synchrotron model well describes the data, we see a particular trend. All the GRBs in the synchrotron sub-sample are in (marginally) fast-cooling regime, as expected from other works \citep{Ravasio2019}.\\ 
Interestingly, the majority of them is in intermediate-cooling regime, which is a specific stage of marginally fast-cooling regime where the two main frequencies $\nu_m$ and  $\nu_c$ are coincident. nearly $60\%$ of GRBs in the \textit{single-bin} Synchrotron sample are in this state, at odds with the $\sim25\%$ of GRBs in \textit{multiple-bins} Synchrotron sample. This suggests that this state is reached more frequently in the brightest pulses.\\

\section{Conclusions} \label{sec:conclusion}
In this work, we perform spectral analysis of GRBs in two samples: the \textit{single-bin} sample (with 74 GRBs, one spectrum with 1s exposure per GRB during their peak) and the \textit{multiple-bins} sample (with 13 bright GRBs, multiple spectra with few seconds exposures per GRB for a total of 125 spectra).\\
In both these samples we fit two main models. The first one is the Band function, which phenomenologically describes GRB spectra without any assumption on the underlying physics. In addition, we used a simple physical model that accounts for power law-distributed relativistic electrons cooling down through synchrotron radiation. \\
We aim to test the validity of the \Ep$-$\Liso\ (Yonetoku) relation using both Band and synchrotron models. While the former was the one historically used to derive GRB empirical relations, the introduction of a synchrotron model, being a viable explanation for GRB prompt spectra, allows to better investigate the nature of this emission and the relative empirical relations.\\
We summarize the main results of this work in the following:
\begin{enumerate}
    \item[a.] We show that most of the GRBs in the \textit{single-bin} sample  are well fitted by the Synchrotron model (58/74). However, there are GRBs whose spectrum cannot be described by a Synchrotron model, preferring instead a Band function (16/74). Nonetheless, GRBs that prefer a Band model, when fitted with a power law at $E < 30$ keV, exhibit a photon index $\alpha_{PL} \lesssim -0.7$, i.e. do not violate the single electron spectrum.\\
    
    \item[b.] It appears that all the GRBs for which we report results from the synchrotron fit are in a fast-cooling regime (\ratio\ $> 1$). The majority of these ones show particular \ratio\ values, where $\nu_m$ (i.e. the spectral peak) is only slightly larger than $\nu_c$ (i.e. the synchrotron break). We refer to these cases as GRBs in an intermediate-cooling state, a particular case of marginally fast-cooling spectra where the break and the peak energies are basically coincident ($1<$ \ratio\ $<3$).\\
    
    \item[c.] We show that the fit of the Band model (single-break spectrum) to a possibly intrinsic synchrotron spectrum (double-break spectrum) leads to systematically lower \Ep\ and the proximity of $E_{b,z}$ to \Ep\ influences the hardness of the spectrum, namely the low-energy spectral index $\alpha$ fitted with the Band model.\\
    
    \item[d.] We observe that GRBs in the \textit{single-bin} sample do not follow the Yonetoku relation \textit{unless} they are in an intermediate-cooling state, i.e.  peak and break energies are extremely close to each other.\\
    \item[e.] We find a novel tight relation between the rest-frame cooling frequency $\nu_{c,z}$ and the isotropic-equivalent luminosity : $$\dfrac{\nu_{c,z}}{100\ keV} = (0.97 \pm 0.14) \times \left(\dfrac{L_{iso}}{10^{52}\ erg/s}\right)^{(0.53 \pm 0.06)}$$
    This relation is found in both \textit{single-bin} and \textit{multiple-bins} samples, with slightly different slopes. \\
\end{enumerate}
The underlying assumption of this work is that prompt emission spectra are produced through synchrotron cooling of power-law distributed particles. Nonetheless, in the \textit{multiple-bins} sample analysis we report synchrotron fit results whenever model comparison does not discard the synchrotron model. This points toward the physical nature of the $\nu_{c,z}-L_{iso}$ relation.\\
We show that, if GRB prompt emission is produced through synchrotron radiation, the fit using a Band function does not always represent a good spectral description. For example, Band fit would return a peak energy estimate $E_{p,z}^{Band}$ in between $h\nu_c$ and $h\nu_m = E_{p,z}^{Sync}$.
This suggests that, for synchrotron spectra, the fundamental physical relation is between \Liso\ and $\nu_{c,z}$. The Yonetoku relation, which was found through Band fits, can be the effect of a single-break model fit to a double-break spectrum. Given the occurrence of GRB spectra in intermediate-cooling regimes (or in general in marginally fast-cooling regimes) $E_{p,z}^{Band}$  provides a proxy of $\nu_{c,z}$. The higher the intrinsic ratio \ratio\ , the higher the scatter of that GRB relative to the Yonetoku relation.\\
This novel interpretation of the Yonetoku relation constitutes a step forward in understanding the physics behind empirical relations and prompt emission. For the prompt synchrotron spectra, we found a new tight relation that connects the cooling regime with the total luminosity of the burst: the less it cools, the brighter it is.\\
This behaviour is counterintuitive. One would expect, in the optically-thin synchrotron framework, more radiative output when particles cool down interacting with stronger magnetic fields.
This new relation can constitute a benchmark in testing GRB prompt emission models, and push forward the possibility of using GRBs in constraining cosmological properties. The recent launch of the SVOM and Einstein Probe missions can provide, in synergies with past observatories, a variety of GRB observations with measured redshift and extend the empirical correlations study with particularly soft burst emitting in the X-rays.

%%%% ACKNOWLEDGEMENTS
\begin{acknowledgements}
The authors thank B. Banerjee, D. Bjørn Malesani, M. Branchesi, A. Celotti, G. Ghirlanda, G. Ghisellini, L. Nava and O.S. Salafia for the fruitful discussions about this work. AM thanks the Cosmic Dawn Center in Copenhagen and the Observatory of Brera in Merate for the hospitality during the development of this project. 
\end{acknowledgements}

%%%% BIBLIOGRAPHY
\bibliographystyle{aa}
\bibliography{references}

\begin{thebibliography}{73}
\expandafter\ifx\csname natexlab\endcsname\relax\def\natexlab#1{#1}\fi

\bibitem[{{Abbott} {et~al.}(2017){Abbott}, {Abbott}, {Abbott}, {Acernese},
  {Ackley}, {Adams}, {Adams}, {Addesso}, {Adhikari}, {Adya}, {Affeldt},
  {Afrough}, {Agarwal}, {Agathos}, {Agatsuma}, {Aggarwal}, {Aguiar}, {Aiello},
  {Ain}, {Ajith}, {Allen}, {Allen}, {Allocca}, {Aloy}, {Altin}, {Amato},
  {Ananyeva}, {Anderson}, {Anderson}, {Angelova}, {Antier}, {Appert}, {Arai},
  {Araya}, {Areeda}, {Arnaud}, {Arun}, {Ascenzi}, {Ashton}, {Ast}, {Aston},
  {Astone}, {Atallah}, {Aufmuth}, {Aulbert}, {AultONeal}, {Austin},
  {Avila-Alvarez}, {Babak}, {Bacon}, {Bader}, {Bae}, {Baker}, {Baldaccini},
  {Ballardin}, {Ballmer}, {Banagiri}, {Barayoga}, {Barclay}, {Barish},
  {Barker}, {Barkett}, {Barone}, {Barr}, {Barsotti}, {Barsuglia}, {Barta},
  {Bartlett}, {Bartos}, {Bassiri}, {Basti}, {Batch}, {Bawaj}, {Bayley},
  {Bazzan}, {B{\'e}csy}, {Beer}, {Bejger}, {Belahcene}, {Bell}, {Berger},
  {Bergmann}, {Bero}, {Berry}, {Bersanetti}, {Bertolini}, {Betzwieser},
  {Bhagwat}, {Bhandare}, {Bilenko}, {Billingsley}, {Billman}, {Birch},
  {Birney}, {Birnholtz}, {Biscans}, {Biscoveanu}, {Bisht}, {Bitossi}, {Biwer},
  {Bizouard}, {Blackburn}, {Blackman}, {Blair}, {Blair}, {Blair}, {Bloemen},
  {Bock}, {Bode}, {Boer}, {Bogaert}, {Bohe}, {Bondu}, {Bonilla}, {Bonnand},
  {Boom}, {Bork}, {Boschi}, {Bose}, {Bossie}, {Bouffanais}, {Bozzi},
  {Bradaschia}, {Brady}, {Branchesi}, {Brau}, {Briant}, {Brillet}, {Brinkmann},
  {Brisson}, {Brockill}, {Broida}, {Brooks}, {Brown}, {Brown}, {Brunett},
  {Buchanan}, {Buikema}, {Bulik}, {Bulten}, {Buonanno}, {Buskulic}, {Buy},
  {Byer}, {Cabero}, {Cadonati}, {Cagnoli}, {Cahillane}, {Calder{\'o}n
  Bustillo}, {Callister}, {Calloni}, {Camp}, {Canepa}, {Canizares}, {Cannon},
  {Cao}, {Cao}, {Capano}, {Capocasa}, {Carbognani}, {Caride}, {Carney},
  {Casanueva Diaz}, {Casentini}, {Caudill}, {Cavagli{\`a}}, {Cavalier},
  {Cavalieri}, {Cella}, {Cepeda}, {Cerd{\'a}-Dur{\'a}n}, {Cerretani},
  {Cesarini}, {Chamberlin}, {Chan}, {Chao}, {Charlton}, {Chase},
  {Chassande-Mottin}, {Chatterjee}, {Chatziioannou}, {Cheeseboro}, {Chen},
  {Chen}, {Chen}, {Cheng}, {Chia}, {Chincarini}, {Chiummo}, {Chmiel}, {Cho},
  {Cho}, {Chow}, {Christensen}, {Chu}, {Chua}, {Chua}, {Chung}, {Chung},
  {Ciani}, {Ciolfi}, {Cirelli}, {Cirone}, {Clara}, {Clark}, {Clearwater},
  {Cleva}, {Cocchieri}, {Coccia}, {Cohadon}, {Cohen}, {Colla}, {Collette},
  {Cominsky}, {Constancio}, {Conti}, {Cooper}, {Corban}, {Corbitt},
  {Cordero-Carri{\'o}n}, {Corley}, {Cornish}, {Corsi}, {Cortese}, {Costa},
  {Coughlin}, {Coughlin}, {Coulon}, {Countryman}, {Couvares}, {Covas}, {Cowan},
  {Coward}, {Cowart}, {Coyne}, {Coyne}, {Creighton}, {Creighton}, {Cripe},
  {Crowder}, {Cullen}, {Cumming}, {Cunningham}, {Cuoco}, {Dal Canton},
  {D{\'a}lya}, {Danilishin}, {D'Antonio}, {Danzmann}, {Dasgupta}, {Da Silva
  Costa}, {Dattilo}, {Dave}, {Davier}, {Davis}, {Daw}, {Day}, {De}, {DeBra},
  {Degallaix}, {De Laurentis}, {Del{\'e}glise}, {Del Pozzo}, {Demos}, {Denker},
  {Dent}, {De Pietri}, {Dergachev}, {De Rosa}, {DeRosa}, {De Rossi}, {DeSalvo},
  {de Varona}, {Devenson}, {Dhurandhar}, {D{\'\i}az}, {Di Fiore}, {Di
  Giovanni}, {Di Girolamo}, {Di Lieto}, {Di Pace}, {Di Palma}, {Di Renzo},
  {Doctor}, {Dolique}, {Donovan}, {Dooley}, {Doravari}, {Dorrington},
  {Douglas}, {Dovale {\'A}lvarez}, {Downes}, {Drago}, {Dreissigacker},
  {Driggers}, {Du}, {Ducrot}, {Dupej}, {Dwyer}, {Edo}, {Edwards}, {Effler},
  {Eggenstein}, {Ehrens}, {Eichholz}, {Eikenberry}, {Eisenstein}, {Essick},
  {Estevez}, {Etienne}, {Etzel}, {Evans}, {Evans}, {Factourovich}, {Fafone},
  {Fair}, {Fairhurst}, {Fan}, {Farinon}, {Farr}, {Farr}, {Fauchon-Jones},
  {Favata}, {Fays}, {Fee}, {Fehrmann}, {Feicht}, {Fejer}, {Fernandez-Galiana},
  {Ferrante}, {Ferreira}, {Ferrini}, {Fidecaro}, {Finstad}, {Fiori},
  {Fiorucci}, {Fishbach}, {Fisher}, {Fitz-Axen}, {Flaminio}, {Fletcher},
  {Fong}, {Font}, {Forsyth}, {Forsyth}, {Fournier}, {Frasca}, {Frasconi},
  {Frei}, {Freise}, {Frey}, {Frey}, {Fries}, {Fritschel}, {Frolov}, {Fulda},
  {Fyffe}, {Gabbard}, {Gadre}, {Gaebel}, {Gair}, {Gammaitoni}, {Ganija},
  {Gaonkar}, {Garcia-Quiros}, {Garufi}, {Gateley}, {Gaudio}, {Gaur},
  {Gayathri}, {Gehrels}, {Gemme}, {Genin}, {Gennai}, {George}, {George},
  {Gergely}, {Germain}, {Ghonge}, {Ghosh}, {Ghosh}, {Ghosh}, {Giaime},
  {Giardina}, {Giazotto}, {Gill}, {Glover}, {Goetz}, {Goetz}, {Gomes},
  {Goncharov}, {Gonz{\'a}lez}, {Gonzalez Castro}, {Gopakumar}, {Gorodetsky},
  {Gossan}, {Gosselin}, {Gouaty}, {Grado}, {Graef}, {Granata}, {Grant}, {Gras},
  {Gray}, {Greco}, {Green}, {Gretarsson}, {Groot}, {Grote}, {Grunewald},
  {Gruning}, {Guidi}, {Guo}, {Gupta}, {Gupta}, {Gushwa}, {Gustafson},
  {Gustafson}, {Halim}, {Hall}, {Hall}, {Hamilton}, {Hammond}, {Haney},
  {Hanke}, {Hanks}, {Hanna}, {Hannam}, {Hannuksela}, {Hanson}, {Hardwick},
  {Harms}, {Harry}, {Harry}, {Hart}, {Haster}, {Haughian}, {Healy}, {Heidmann},
  {Heintze}, {Heitmann}, {Hello}, {Hemming}, {Hendry}, {Heng}, {Hennig},
  {Heptonstall}, {Heurs}, {Hild}, {Hinderer}, {Hoak}, {Hofman}, {Holt}, {Holz},
  {Hopkins}, {Horst}, {Hough}, {Houston}, {Howell}, {Hreibi}, {Hu}, {Huerta},
  {Huet}, {Hughey}, {Husa}, {Huttner}, {Huynh-Dinh}, {Indik}, {Inta}, {Intini},
  {Isa}, {Isac}, {Isi}, {Iyer}, {Izumi}, {Jacqmin}, {Jani}, {Jaranowski},
  {Jawahar}, {Jim{\'e}nez-Forteza}, {Johnson}, {Johnson-McDaniel}, {Jones},
  {Jones}, {Jonker}, {Ju}, {Junker}, {Kalaghatgi}, {Kalogera}, {Kamai},
  {Kandhasamy}, {Kang}, {Kanner}, {Kapadia}, {Karki}, {Karvinen}, {Kasprzack},
  {Kastaun}, {Katolik}, {Katsavounidis}, {Katzman}, {Kaufer}, {Kawabe},
  {K{\'e}f{\'e}lian}, {Keitel}, {Kemball}, {Kennedy}, {Kent}, {Key}, {Khalili},
  {Khan}, {Khan}, {Khan}, {Khazanov}, {Kijbunchoo}, {Kim}, {Kim}, {Kim}, {Kim},
  {Kim}, {Kim}, {Kimbrell}, {King}, {King}, {Kinley-Hanlon}, {Kirchhoff},
  {Kissel}, {Kleybolte}, {Klimenko}, {Knowles}, {Koch}, {Koehlenbeck}, {Koley},
  {Kondrashov}, {Kontos}, {Korobko}, {Korth}, {Kowalska}, {Kozak},
  {Kr{\"a}mer}, {Kringel}, {Krishnan}, {Kr{\'o}lak}, {Kuehn}, {Kumar}, {Kumar},
  {Kumar}, {Kuo}, {Kutynia}, {Kwang}, {Lackey}, {Lai}, {Landry}, {Lang},
  {Lange}, {Lantz}, {Lanza}, {Lartaux-Vollard}, {Lasky}, {Laxen}, {Lazzarini},
  {Lazzaro}, {Leaci}, {Leavey}, {Lee}, {Lee}, {Lee}, {Lee}, {Lee}, {Lehmann},
  {Lenon}, {Leonardi}, {Leroy}, {Letendre}, {Levin}, {Li}, {Linker},
  {Littenberg}, {Liu}, {Lo}, {Lockerbie}, {London}, {Lord}, {Lorenzini},
  {Loriette}, {Lormand}, {Losurdo}, {Lough}, {Lousto}, {Lovelace}, {L{\"u}ck},
  {Lumaca}, {Lundgren}, {Lynch}, {Ma}, {Macas}, {Macfoy}, {Machenschalk},
  {MacInnis}, {Macleod}, {Maga{\~n}a Hernandez}, {Maga{\~n}a-Sandoval},
  {Maga{\~n}a Zertuche}, {Magee}, {Majorana}, {Maksimovic}, {Man}, {Mandic},
  {Mangano}, {Mansell}, {Manske}, {Mantovani}, {Marchesoni}, {Marion},
  {M{\'a}rka}, {M{\'a}rka}, {Markakis}, {Markosyan}, {Markowitz}, {Maros},
  {Marquina}, {Martelli}, {Martellini}, {Martin}, {Martin}, {Martynov},
  {Mason}, {Massera}, {Masserot}, {Massinger}, {Masso-Reid}, {Mastrogiovanni},
  {Matas}, {Matichard}, {Matone}, {Mavalvala}, {Mazumder}, {McCarthy},
  {McClelland}, {McCormick}, {McCuller}, {McGuire}, {McIntyre}, {McIver},
  {McManus}, {McNeill}, {McRae}, {McWilliams}, {Meacher}, {Meadors}, {Mehmet},
  {Meidam}, {Mejuto-Villa}, {Melatos}, {Mendell}, {Mercer}, {Merilh},
  {Merzougui}, {Meshkov}, {Messenger}, {Messick}, {Metzdorff}, {Meyers},
  {Miao}, {Michel}, {Middleton}, {Mikhailov}, {Milano}, {Miller}, {Miller},
  {Miller}, {Millhouse}, {Milovich-Goff}, {Minazzoli}, {Minenkov}, {Ming},
  {Mishra}, {Mitra}, {Mitrofanov}, {Mitselmakher}, {Mittleman}, {Moffa},
  {Moggi}, {Mogushi}, {Mohan}, {Mohapatra}, {Montani}, {Moore}, {Moraru},
  {Moreno}, {Morriss}, {Mours}, {Mow-Lowry}, {Mueller}, {Muir}, {Mukherjee},
  {Mukherjee}, {Mukherjee}, {Mukund}, {Mullavey}, {Munch}, {Mu{\~n}iz},
  {Muratore}, {Murray}, {Napier}, {Nardecchia}, {Naticchioni}, {Nayak},
  {Neilson}, {Nelemans}, {Nelson}, {Nery}, {Neunzert}, {Nevin}, {Newport},
  {Newton}, {Ng}, {Nguyen}, {Nichols}, {Nielsen}, {Nissanke}, {Nitz}, {Noack},
  {Nocera}, {Nolting}, {North}, {Nuttall}, {Oberling}, {O'Dea}, {Ogin}, {Oh},
  {Oh}, {Ohme}, {Okada}, {Oliver}, {Oppermann}, {Oram}, {O'Reilly}, {Ormiston},
  {Ortega}, {O'Shaughnessy}, {Ossokine}, {Ottaway}, {Overmier}, {Owen}, {Pace},
  {Page}, {Page}, {Pai}, {Pai}, {Palamos}, {Palashov}, {Palomba}, {Pal-Singh},
  {Pan}, {Pan}, {Pang}, {Pang}, {Pankow}, {Pannarale}, {Pant}, {Paoletti},
  {Paoli}, {Papa}, {Parida}, {Parker}, {Pascucci}, {Pasqualetti},
  {Passaquieti}, {Passuello}, {Patil}, {Patricelli}, {Pearlstone}, {Pedraza},
  {Pedurand}, {Pekowsky}, {Pele}, {Penn}, {Perez}, {Perreca}, {Perri},
  {Pfeiffer}, {Phelps}, {Piccinni}, {Pichot}, {Piergiovanni}, {Pierro},
  {Pillant}, {Pinard}, {Pinto}, {Pirello}, {Pitkin}, {Poe}, {Poggiani},
  {Popolizio}, {Porter}, {Post}, {Powell}, {Prasad}, {Pratt}, {Pratten},
  {Predoi}, {Prestegard}, {Prijatelj}, {Principe}, {Privitera}, {Prodi},
  {Prokhorov}, {Puncken}, {Punturo}, {Puppo}, {P{\"u}rrer}, {Qi}, {Quetschke},
  {Quintero}, {Quitzow-James}, {Raab}, {Rabeling}, {Radkins}, {Raffai}, {Raja},
  {Rajan}, {Rajbhandari}, {Rakhmanov}, {Ramirez}, {Ramos-Buades}, {Rapagnani},
  {Raymond}, {Razzano}, {Read}, {Regimbau}, {Rei}, {Reid}, {Reitze}, {Ren},
  {Reyes}, {Ricci}, {Ricker}, {Rieger}, {Riles}, {Rizzo}, {Robertson}, {Robie},
  {Robinet}, {Rocchi}, {Rolland}, {Rollins}, {Roma}, {Romano}, {Romel},
  {Romie}, {Rosi{\'n}ska}, {Ross}, {Rowan}, {R{\"u}diger}, {Ruggi}, {Rutins},
  {Ryan}, {Sachdev}, {Sadecki}, {Sadeghian}, {Sakellariadou}, {Salconi},
  {Saleem}, {Salemi}, {Samajdar}, {Sammut}, {Sampson}, {Sanchez}, {Sanchez},
  {Sanchis-Gual}, {Sandberg}, {Sanders}, {Sassolas}, {Sathyaprakash},
  {Saulson}, {Sauter}, {Savage}, {Sawadsky}, {Schale}, {Scheel}, {Scheuer},
  {Schmidt}, {Schmidt}, {Schnabel}, {Schofield}, {Sch{\"o}nbeck}, {Schreiber},
  {Schuette}, {Schulte}, {Schutz}, {Schwalbe}, {Scott}, {Scott}, {Seidel},
  {Sellers}, {Sengupta}, {Sentenac}, {Sequino}, {Sergeev}, {Shaddock},
  {Shaffer}, {Shah}, {Shahriar}, {Shaner}, {Shao}, {Shapiro}, {Shawhan},
  {Sheperd}, {Shoemaker}, {Shoemaker}, {Siellez}, {Siemens}, {Sieniawska},
  {Sigg}, {Silva}, {Singer}, {Singh}, {Singhal}, {Sintes}, {Slagmolen},
  {Smith}, {Smith}, {Smith}, {Somala}, {Son}, {Sonnenberg}, {Sorazu},
  {Sorrentino}, {Souradeep}, {Spencer}, {Srivastava}, {Staats}, {Staley},
  {Steinke}, {Steinlechner}, {Steinlechner}, {Steinmeyer}, {Stevenson},
  {Stone}, {Stops}, {Strain}, {Stratta}, {Strigin}, {Strunk}, {Sturani},
  {Stuver}, {Summerscales}, {Sun}, {Sunil}, {Suresh}, {Sutton}, {Swinkels},
  {Szczepa{\'n}czyk}, {Tacca}, {Tait}, {Talbot}, {Talukder}, {Tanner},
  {T{\'a}pai}, {Taracchini}, {Tasson}, {Taylor}, {Taylor}, {Tewari}, {Theeg},
  {Thies}, {Thomas}, {Thomas}, {Thomas}, {Thorne}, {Thorne}, {Thrane},
  {Tiwari}, {Tiwari}, {Tokmakov}, {Toland}, {Tonelli}, {Tornasi},
  {Torres-Forn{\'e}}, {Torrie}, {T{\"o}yr{\"a}}, {Travasso}, {Traylor},
  {Trinastic}, {Tringali}, {Trozzo}, {Tsang}, {Tse}, {Tso}, {Tsukada}, {Tsuna},
  {Tuyenbayev}, {Ueno}, {Ugolini}, {Unnikrishnan}, {Urban}, {Usman},
  {Vahlbruch}, {Vajente}, {Valdes}, {van Bakel}, {van Beuzekom}, {van den
  Brand}, {Van Den Broeck}, {Vander-Hyde}, {van der Schaaf}, {van Heijningen},
  {van Veggel}, {Vardaro}, {Varma}, {Vass}, {Vas{\'u}th}, {Vecchio},
  {Vedovato}, {Veitch}, {Veitch}, {Venkateswara}, {Venugopalan}, {Verkindt},
  {Vetrano}, {Vicer{\'e}}, {Viets}, {Vinciguerra}, {Vine}, {Vinet}, {Vitale},
  {Vo}, {Vocca}, {Vorvick}, {Vyatchanin}, {Wade}, {Wade}, {Wade}, {Walet},
  {Walker}, {Wallace}, {Walsh}, {Wang}, {Wang}, {Wang}, {Wang}, {Wang}, {Ward},
  {Warner}, {Was}, {Watchi}, {Weaver}, {Wei}, {Weinert}, {Weinstein}, {Weiss},
  {Wen}, {Wessel}, {We{\ss}els}, {Westerweck}, {Westphal}, {Wette}, {Whelan},
  {Whitcomb}, {Whiting}, {Whittle}, {Wilken}, {Williams}, {Williams},
  {Williamson}, {Willis}, {Willke}, {Wimmer}, {Winkler}, {Wipf}, {Wittel},
  {Woan}, {Woehler}, {Wofford}, {Wong}, {Worden}, {Wright}, {Wu}, {Wysocki},
  {Xiao}, {Yamamoto}, {Yancey}, {Yang}, {Yap}, {Yazback}, {Yu}, {Yu}, {Yvert},
  {Zadro{\.z}ny}, {Zanolin}, {Zelenova}, {Zendri}, {Zevin}, {Zhang}, {Zhang},
  {Zhang}, {Zhang}, {Zhao}, {Zhou}, {Zhou}, {Zhu}, {Zhu}, {Zimmerman},
  {Zucker}, {Zweizig}, {(LIGO Scientific Collaboration}, {Virgo Collaboration},
  {Burns}, {Veres}, {Kocevski}, {Racusin}, {Goldstein}, {Connaughton},
  {Briggs}, {Blackburn}, {Hamburg}, {Hui}, {von Kienlin}, {McEnery}, {Preece},
  {Wilson-Hodge}, {Bissaldi}, {Cleveland}, {Gibby}, {Giles}, {Kippen},
  {McBreen}, {Meegan}, {Paciesas}, {Poolakkil}, {Roberts}, {Stanbro},
  {Gamma-ray Burst Monitor}, {Savchenko}, {Ferrigno}, {Kuulkers}, {Bazzano},
  {Bozzo}, {Brandt}, {Chenevez}, {Courvoisier}, {Diehl}, {Domingo}, {Hanlon},
  {Jourdain}, {Laurent}, {Lebrun}, {Lutovinov}, {Mereghetti}, {Natalucci},
  {Rodi}, {Roques}, {Sunyaev}, {Ubertini}, \& {(INTEGRAL}}]{Abbott2017}
{Abbott}, B.~P., {Abbott}, R., {Abbott}, T.~D., {et~al.} 2017, \apjl, 848, L13

\bibitem[{{Abdo} {et~al.}(2009){Abdo}, {Ackermann}, {Ajello}, {Asano},
  {Atwood}, {Axelsson}, {Baldini}, {Ballet}, {Barbiellini}, {Baring},
  {Bastieri}, {Bechtol}, {Bellazzini}, {Berenji}, {Bhat}, {Bissaldi},
  {Blandford}, {Bloom}, {Bonamente}, {Borgland}, {Bouvier}, {Bregeon}, {Brez},
  {Briggs}, {Brigida}, {Bruel}, {Burgess}, {Burrows}, {Buson}, {Caliandro},
  {Cameron}, {Caraveo}, {Casandjian}, {Cecchi}, {{\c{C}}elik}, {Chekhtman},
  {Cheung}, {Chiang}, {Ciprini}, {Claus}, {Cohen-Tanugi}, {Cominsky},
  {Connaughton}, {Conrad}, {Cutini}, {d'Elia}, {Dermer}, {de Angelis}, {de
  Palma}, {Digel}, {Dingus}, {Silva}, {Drell}, {Dubois}, {Dumora}, {Farnier},
  {Favuzzi}, {Fegan}, {Finke}, {Fishman}, {Focke}, {Fortin}, {Frailis},
  {Fukazawa}, {Funk}, {Fusco}, {Gargano}, {Gehrels}, {Germani}, {Giavitto},
  {Giebels}, {Giglietto}, {Giordano}, {Glanzman}, {Godfrey}, {Goldstein},
  {Granot}, {Greiner}, {Grenier}, {Grove}, {Guillemot}, {Guiriec}, {Hanabata},
  {Harding}, {Hayashida}, {Hays}, {Horan}, {Hughes}, {Jackson},
  {J{\'o}hannesson}, {Johnson}, {Johnson}, {Johnson}, {Kamae}, {Katagiri},
  {Kataoka}, {Kawai}, {Kerr}, {Kippen}, {Kn{\"o}dlseder}, {Kocevski}, {Komin},
  {Kouveliotou}, {Kuss}, {Lande}, {Latronico}, {Lemoine-Goumard}, {Longo},
  {Loparco}, {Lott}, {Lovellette}, {Lubrano}, {Madejski}, {Makeev},
  {Mazziotta}, {McBreen}, {McEnery}, {McGlynn}, {Meegan}, {M{\'e}sz{\'a}ros},
  {Meurer}, {Michelson}, {Mitthumsiri}, {Mizuno}, {Moiseev}, {Monte},
  {Monzani}, {Moretti}, {Morselli}, {Moskalenko}, {Murgia}, {Nakamori},
  {Nolan}, {Norris}, {Nuss}, {Ohno}, {Ohsugi}, {Omodei}, {Orlando}, {Ormes},
  {Paciesas}, {Paneque}, {Panetta}, {Pelassa}, {Pepe}, {Pesce-Rollins},
  {Petrosian}, {Piron}, {Porter}, {Preece}, {Rain{\`o}}, {Rando}, {Rau},
  {Razzano}, {Razzaque}, {Reimer}, {Reimer}, {Reposeur}, {Ritz}, {Rochester},
  {Rodriguez}, {Roming}, {Roth}, {Ryde}, {Sadrozinski}, {Sanchez}, {Sander},
  {Saz Parkinson}, {Scargle}, {Schalk}, {Sgr{\`o}}, {Siskind}, {Smith},
  {Spinelli}, {Stamatikos}, {Stecker}, {Stratta}, {Strickman}, {Suson},
  {Swenson}, {Tajima}, {Takahashi}, {Tanaka}, {Thayer}, {Thayer}, {Thompson},
  {Tibaldo}, {Torres}, {Tosti}, {Tramacere}, {Uchiyama}, {Uehara}, {Usher},
  {van der Horst}, {Vasileiou}, {Vilchez}, {Vitale}, {von Kienlin}, {Waite},
  {Wang}, {Wilson-Hodge}, {Winer}, {Wood}, {Yamazaki}, {Ylinen}, \&
  {Ziegler}}]{Fermi2009}
{Abdo}, A.~A., {Ackermann}, M., {Ajello}, M., {et~al.} 2009, \apjl, 706, L138

\bibitem[{{Ajello} {et~al.}(2020){Ajello}, {Arimoto}, {Axelsson}, {Baldini},
  {Barbiellini}, {Bastieri}, {Bellazzini}, {Berretta}, {Bissaldi}, {Blandford},
  {Bonino}, {Bottacini}, {Bregeon}, {Bruel}, {Buehler}, {Burns}, {Buson},
  {Cameron}, {Caputo}, {Caraveo}, {Cavazzuti}, {Chen}, {Chiaro}, {Ciprini},
  {Cohen-Tanugi}, {Costantin}, {Cutini}, {D'Ammando}, {DeKlotz}, {de la Torre
  Luque}, {de Palma}, {Desai}, {Di Lalla}, {Di Venere}, {Fana Dirirsa},
  {Fegan}, {Franckowiak}, {Fukazawa}, {Funk}, {Fusco}, {Gargano}, {Gasparrini},
  {Giglietto}, {Gill}, {Giordano}, {Giroletti}, {Granot}, {Green}, {Grenier},
  {Grondin}, {Guiriec}, {Hays}, {Horan}, {J{\'o}hannesson}, {Kocevski},
  {Kovac'evic'}, {Kuss}, {Larsson}, {Latronico}, {Lemoine-Goumard}, {Li},
  {Liodakis}, {Longo}, {Loparco}, {Lovellette}, {Lubrano}, {Maldera},
  {Malyshev}, {Manfreda}, {Mart{\'\i}-Devesa}, {Mazziotta}, {McEnery}, {Mereu},
  {Meyer}, {Michelson}, {Mitthumsiri}, {Mizuno}, {Monzani}, {Moretti},
  {Morselli}, {Moskalenko}, {Negro}, {Nuss}, {Omodei}, {Orienti}, {Orlando},
  {Palatiello}, {Paliya}, {Paneque}, {Pei}, {Persic}, {Pesce-Rollins},
  {Petrosian}, {Piron}, {Poon}, {Porter}, {Principe}, {Racusin}, {Rain{\`o}},
  {Rando}, {Rani}, {Razzano}, {Razzaque}, {Reimer}, {Reimer}, {Ryde}, {Saz
  Parkinson}, {Serini}, {Sgr{\`o}}, {Siskind}, {Spandre}, {Spinelli}, {Tajima},
  {Takagi}, {Takahashi}, {Tak}, {Thayer}, {Thompson}, {Torres}, {Troja},
  {Valverde}, {Van Klaveren}, {Wood}, {Yassine}, {Zaharijas}, {Mailyan},
  {Bhat}, {Briggs}, {Cleveland}, {Giles}, {Goldstein}, {Hui}, {Malacaria},
  {Preece}, {Roberts}, {Veres}, {Wilson-Hodge}, {Kienlin}, {Cenko}, {O'Brien},
  {Beardmore}, {Lien}, {Osborne}, {Tohuvavohu}, {D'Elia}, {D'A{\`\i}}, {Perri},
  {Gropp}, {Klingler}, {Capalbi}, {Tagliaferri}, {Stamatikos}, \& {De
  Pasquale}}]{Ajello2020}
{Ajello}, M., {Arimoto}, M., {Axelsson}, M., {et~al.} 2020, \apj, 890, 9

\bibitem[{{Amati} {et~al.}(2002){Amati}, {Frontera}, {Tavani}, {in't Zand},
  {Antonelli}, {Costa}, {Feroci}, {Guidorzi}, {Heise}, {Masetti}, {Montanari},
  {Nicastro}, {Palazzi}, {Pian}, {Piro}, \& {Soffitta}}]{Amati2002}
{Amati}, L., {Frontera}, F., {Tavani}, M., {et~al.} 2002, \aap, 390, 81

\bibitem[{{Arnaud}(1996)}]{Arnaud1996}
{Arnaud}, K.~A. 1996, in Astronomical Society of the Pacific Conference Series,
  Vol. 101, Astronomical Data Analysis Software and Systems V, ed. G.~H.
  {Jacoby} \& J.~{Barnes}, 17

\bibitem[{{Band} {et~al.}(1993){Band}, {Matteson}, {Ford}, {Schaefer},
  {Palmer}, {Teegarden}, {Cline}, {Briggs}, {Paciesas}, {Pendleton}, {Fishman},
  {Kouveliotou}, {Meegan}, {Wilson}, \& {Lestrade}}]{Band1993}
{Band}, D., {Matteson}, J., {Ford}, L., {et~al.} 1993, \apj, 413, 281

\bibitem[{{Band} \& {Preece}(2005)}]{Band&Preece2005}
{Band}, D.~L. \& {Preece}, R.~D. 2005, \apj, 627, 319

\bibitem[{{Berger}(2014)}]{Berger2014}
{Berger}, E. 2014, \araa, 52, 43

\bibitem[{{Bhat} {et~al.}(2012){Bhat}, {Briggs}, {Connaughton}, {Kouveliotou},
  {van der Horst}, {Paciesas}, {Meegan}, {Bissaldi}, {Burgess}, {Chaplin},
  {Diehl}, {Fishman}, {Fitzpatrick}, {Foley}, {Gibby}, {Giles}, {Goldstein},
  {Greiner}, {Gruber}, {Guiriec}, {von Kienlin}, {Kippen}, {McBreen}, {Preece},
  {Rau}, {Tierney}, \& {Wilson-Hodge}}]{Bhat2012}
{Bhat}, P.~N., {Briggs}, M.~S., {Connaughton}, V., {et~al.} 2012, \apj, 744,
  141

\bibitem[{{Bo{\v{s}}njak} {et~al.}(2009){Bo{\v{s}}njak}, {Daigne}, \&
  {Dubus}}]{Bosnjak2009}
{Bo{\v{s}}njak}, {\v{Z}}., {Daigne}, F., \& {Dubus}, G. 2009, \aap, 498, 677

\bibitem[{{Buchner}(2016)}]{Buchner2016}
{Buchner}, J. 2016, {BXA: Bayesian X-ray Analysis}, Astrophysics Source Code
  Library, record ascl:1610.011

\bibitem[{{Burgess} {et~al.}(2020){Burgess}, {B{\'e}gu{\'e}}, {Greiner},
  {Giannios}, {Bacelj}, \& {Berlato}}]{Burgess2020}
{Burgess}, J.~M., {B{\'e}gu{\'e}}, D., {Greiner}, J., {et~al.} 2020, Nature
  Astronomy, 4, 174

\bibitem[{{Burgess} {et~al.}(2015){Burgess}, {Ryde}, \& {Yu}}]{Burgess2015}
{Burgess}, J.~M., {Ryde}, F., \& {Yu}, H.-F. 2015, \mnras, 451, 1511

\bibitem[{{Butler} {et~al.}(2007){Butler}, {Kocevski}, {Bloom}, \&
  {Curtis}}]{Butler2007}
{Butler}, N.~R., {Kocevski}, D., {Bloom}, J.~S., \& {Curtis}, J.~L. 2007, \apj,
  671, 656

\bibitem[{{D'Agostini}(2005)}]{D'Agostini2005}
{D'Agostini}, G. 2005, arXiv e-prints, physics/0511182

\bibitem[{{Daigne} \& {Bo{\v{s}}njak}(2024)}]{Daigne2024}
{Daigne}, F. \& {Bo{\v{s}}njak}, {\v{Z}}. 2024, arXiv e-prints,
  arXiv:2407.04023

\bibitem[{{Daigne} {et~al.}(2011){Daigne}, {Bo{\v{s}}njak}, \&
  {Dubus}}]{Daigne2011}
{Daigne}, F., {Bo{\v{s}}njak}, {\v{Z}}., \& {Dubus}, G. 2011, \aap, 526, A110

\bibitem[{{Derishev}(2007)}]{Derishev2007}
{Derishev}, E.~V. 2007, \apss, 309, 157

\bibitem[{{Derishev}(2009)}]{Derishev2009}
{Derishev}, E.~V. 2009, International Journal of Modern Physics D, 18, 1523

\bibitem[{{Drenkhahn} \& {Spruit}(2002)}]{Drenkhahn&Spruit2002}
{Drenkhahn}, G. \& {Spruit}, H.~C. 2002, \aap, 391, 1141

\bibitem[{{Eichler} {et~al.}(1989){Eichler}, {Livio}, {Piran}, \&
  {Schramm}}]{Eichler1989}
{Eichler}, D., {Livio}, M., {Piran}, T., \& {Schramm}, D.~N. 1989, \nat, 340,
  126

\bibitem[{{Foreman-Mackey} {et~al.}(2013){Foreman-Mackey}, {Hogg}, {Lang}, \&
  {Goodman}}]{emcee}
{Foreman-Mackey}, D., {Hogg}, D.~W., {Lang}, D., \& {Goodman}, J. 2013, \pasp,
  125, 306

\bibitem[{{Ghirlanda} {et~al.}(2011{\natexlab{a}}){Ghirlanda}, {Ghisellini}, \&
  {Nava}}]{Ghirlanda2011a}
{Ghirlanda}, G., {Ghisellini}, G., \& {Nava}, L. 2011{\natexlab{a}}, \mnras,
  418, L109

\bibitem[{{Ghirlanda} {et~al.}(2011{\natexlab{b}}){Ghirlanda}, {Ghisellini},
  {Nava}, \& {Burlon}}]{Ghirlanda2011b}
{Ghirlanda}, G., {Ghisellini}, G., {Nava}, L., \& {Burlon}, D.
  2011{\natexlab{b}}, \mnras, 410, L47

\bibitem[{{Ghirlanda} {et~al.}(2010){Ghirlanda}, {Nava}, \&
  {Ghisellini}}]{Ghirlanda2010}
{Ghirlanda}, G., {Nava}, L., \& {Ghisellini}, G. 2010, \aap, 511, A43

\bibitem[{{Ghirlanda} {et~al.}(2009){Ghirlanda}, {Nava}, {Ghisellini},
  {Celotti}, \& {Firmani}}]{Ghirlanda2009}
{Ghirlanda}, G., {Nava}, L., {Ghisellini}, G., {Celotti}, A., \& {Firmani}, C.
  2009, \aap, 496, 585

\bibitem[{{Ghirlanda} {et~al.}(2008){Ghirlanda}, {Nava}, {Ghisellini},
  {Firmani}, \& {Cabrera}}]{Ghirlanda2008}
{Ghirlanda}, G., {Nava}, L., {Ghisellini}, G., {Firmani}, C., \& {Cabrera},
  J.~I. 2008, \mnras, 387, 319

\bibitem[{{Ghisellini} \& {Celotti}(1999)}]{Ghisellini&Celotti1999}
{Ghisellini}, G. \& {Celotti}, A. 1999, \apjl, 511, L93

\bibitem[{{Golenetskii} {et~al.}(1983){Golenetskii}, {Mazets}, {Aptekar}, \&
  {Ilinskii}}]{Golenetskii1983}
{Golenetskii}, S.~V., {Mazets}, E.~P., {Aptekar}, R.~L., \& {Ilinskii}, V.~N.
  1983, \nat, 306, 451

\bibitem[{{Gompertz} {et~al.}(2023){Gompertz}, {Ravasio}, {Nicholl}, {Levan},
  {Metzger}, {Oates}, {Lamb}, {Fong}, {Malesani}, {Rastinejad}, {Tanvir},
  {Evans}, {Jonker}, {Page}, \& {Pe'er}}]{Gompertz2023}
{Gompertz}, B.~P., {Ravasio}, M.~E., {Nicholl}, M., {et~al.} 2023, Nature
  Astronomy, 7, 67

\bibitem[{{Gruber} {et~al.}(2014){Gruber}, {Goldstein}, {Weller von Ahlefeld},
  {Narayana Bhat}, {Bissaldi}, {Briggs}, {Byrne}, {Cleveland}, {Connaughton},
  {Diehl}, {Fishman}, {Fitzpatrick}, {Foley}, {Gibby}, {Giles}, {Greiner},
  {Guiriec}, {van der Horst}, {von Kienlin}, {Kouveliotou}, {Layden}, {Lin},
  {Meegan}, {McGlynn}, {Paciesas}, {Pelassa}, {Preece}, {Rau}, {Wilson-Hodge},
  {Xiong}, {Younes}, \& {Yu}}]{Gruber2014}
{Gruber}, D., {Goldstein}, A., {Weller von Ahlefeld}, V., {et~al.} 2014, \apjs,
  211, 12

\bibitem[{{Guiriec} {et~al.}(2013){Guiriec}, {Daigne}, {Hasco{\"e}t},
  {Vianello}, {Ryde}, {Mochkovitch}, {Kouveliotou}, {Xiong}, {Bhat}, {Foley},
  {Gruber}, {Burgess}, {McGlynn}, {McEnery}, \& {Gehrels}}]{Guiriec2013}
{Guiriec}, S., {Daigne}, F., {Hasco{\"e}t}, R., {et~al.} 2013, \apj, 770, 32

\bibitem[{{Guiriec} {et~al.}(2015{\natexlab{a}}){Guiriec}, {Kouveliotou},
  {Daigne}, {Zhang}, {Hasco{\"e}t}, {Nemmen}, {Thompson}, {Bhat}, {Gehrels},
  {Gonzalez}, {Kaneko}, {McEnery}, {Mochkovitch}, {Racusin}, {Ryde}, {Sacahui},
  \& {{\"U}nsal}}]{Guiriec2015}
{Guiriec}, S., {Kouveliotou}, C., {Daigne}, F., {et~al.} 2015{\natexlab{a}},
  \apj, 807, 148

\bibitem[{{Guiriec} {et~al.}(2015{\natexlab{b}}){Guiriec}, {Mochkovitch},
  {Piran}, {Daigne}, {Kouveliotou}, {Racusin}, {Gehrels}, \&
  {McEnery}}]{Guiriec2015b}
{Guiriec}, S., {Mochkovitch}, R., {Piran}, T., {et~al.} 2015{\natexlab{b}},
  \apj, 814, 10

\bibitem[{{Kaneko} {et~al.}(2006){Kaneko}, {Preece}, {Briggs}, {Paciesas},
  {Meegan}, \& {Band}}]{Kaneko2006}
{Kaneko}, Y., {Preece}, R.~D., {Briggs}, M.~S., {et~al.} 2006, \apjs, 166, 298

\bibitem[{Kass \& Raftery(1995)}]{kass_raftery95}
Kass, R.~E. \& Raftery, A.~E. 1995, Journal of the American Statistical
  Association, 90, 773

\bibitem[{{Lazzati} {et~al.}(2000){Lazzati}, {Ghisellini}, {Celotti}, \&
  {Rees}}]{Lazzati2000}
{Lazzati}, D., {Ghisellini}, G., {Celotti}, A., \& {Rees}, M.~J. 2000, \apjl,
  529, L17

\bibitem[{{Lu} {et~al.}(2012){Lu}, {Wei}, {Liang}, {Zhang}, {L{\"u}}, {L{\"u}},
  {Lei}, \& {Zhang}}]{Lu2012}
{Lu}, R.-J., {Wei}, J.-J., {Liang}, E.-W., {et~al.} 2012, \apj, 756, 112

\bibitem[{{Medvedev}(2000)}]{Medvedev2000}
{Medvedev}, M.~V. 2000, \apj, 540, 704

\bibitem[{{Meegan} {et~al.}(2009){Meegan}, {Lichti}, {Bhat}, {Bissaldi},
  {Briggs}, {Connaughton}, {Diehl}, {Fishman}, {Greiner}, {Hoover}, {van der
  Horst}, {von Kienlin}, {Kippen}, {Kouveliotou}, {McBreen}, {Paciesas},
  {Preece}, {Steinle}, {Wallace}, {Wilson}, \& {Wilson-Hodge}}]{Meegan2009}
{Meegan}, C., {Lichti}, G., {Bhat}, P.~N., {et~al.} 2009, \apj, 702, 791

\bibitem[{{M{\'e}sz{\'a}ros} \& {Rees}(2000)}]{Meszaros&Rees2000}
{M{\'e}sz{\'a}ros}, P. \& {Rees}, M.~J. 2000, \apj, 530, 292

\bibitem[{{Minaev} \& {Pozanenko}(2020)}]{Minaev2020}
{Minaev}, P.~Y. \& {Pozanenko}, A.~S. 2020, \mnras, 492, 1919

\bibitem[{{Nakar} \& {Piran}(2005)}]{Nakar&Piran2005}
{Nakar}, E. \& {Piran}, T. 2005, \mnras, 360, L73

\bibitem[{{Narayan} {et~al.}(1992){Narayan}, {Paczynski}, \&
  {Piran}}]{Narayan&Paczynski1992}
{Narayan}, R., {Paczynski}, B., \& {Piran}, T. 1992, \apjl, 395, L83

\bibitem[{{Nava} {et~al.}(2008){Nava}, {Ghirlanda}, {Ghisellini}, \&
  {Firmani}}]{Nava2008}
{Nava}, L., {Ghirlanda}, G., {Ghisellini}, G., \& {Firmani}, C. 2008, \mnras,
  391, 639

\bibitem[{{Nava} {et~al.}(2012){Nava}, {Salvaterra}, {Ghirlanda}, {Ghisellini},
  {Campana}, {Covino}, {Cusumano}, {D'Avanzo}, {D'Elia}, {Fugazza}, {Melandri},
  {Sbarufatti}, {Vergani}, \& {Tagliaferri}}]{Nava2012}
{Nava}, L., {Salvaterra}, R., {Ghirlanda}, G., {et~al.} 2012, \mnras, 421, 1256

\bibitem[{{Oganesyan} {et~al.}(2017){Oganesyan}, {Nava}, {Ghirlanda}, \&
  {Celotti}}]{Oganesyan2017}
{Oganesyan}, G., {Nava}, L., {Ghirlanda}, G., \& {Celotti}, A. 2017, \apj, 846,
  137

\bibitem[{{Oganesyan} {et~al.}(2018){Oganesyan}, {Nava}, {Ghirlanda}, \&
  {Celotti}}]{Oganesyan2018}
{Oganesyan}, G., {Nava}, L., {Ghirlanda}, G., \& {Celotti}, A. 2018, \aap, 616,
  A138

\bibitem[{{Oganesyan} {et~al.}(2019){Oganesyan}, {Nava}, {Ghirlanda},
  {Melandri}, \& {Celotti}}]{Oganesyan2019}
{Oganesyan}, G., {Nava}, L., {Ghirlanda}, G., {Melandri}, A., \& {Celotti}, A.
  2019, \aap, 628, A59

\bibitem[{{Paczynski}(1986)}]{Paczynski1986}
{Paczynski}, B. 1986, \apjl, 308, L43

\bibitem[{{Parsotan} \& {Ito}(2022)}]{Parsotan&Ito2022}
{Parsotan}, T. \& {Ito}, H. 2022, Universe, 8, 310

\bibitem[{{Pe'er}(2008)}]{Pe'er2008}
{Pe'er}, A. 2008, \apj, 682, 463

\bibitem[{{Pe'er} \& {Zhang}(2006)}]{Pe'er2006}
{Pe'er}, A. \& {Zhang}, B. 2006, \apj, 653, 454

\bibitem[{{Preece} {et~al.}(1998){Preece}, {Briggs}, {Mallozzi}, {Pendleton},
  {Paciesas}, \& {Band}}]{Preece1998}
{Preece}, R.~D., {Briggs}, M.~S., {Mallozzi}, R.~S., {et~al.} 1998, \apjl, 506,
  L23

\bibitem[{{Ravasio} {et~al.}(2019){Ravasio}, {Ghirlanda}, {Nava}, \&
  {Ghisellini}}]{Ravasio2019}
{Ravasio}, M.~E., {Ghirlanda}, G., {Nava}, L., \& {Ghisellini}, G. 2019, \aap,
  625, A60

\bibitem[{{Rees} \& {Meszaros}(1994)}]{Rees&Meszaros1994}
{Rees}, M.~J. \& {Meszaros}, P. 1994, \apjl, 430, L93

\bibitem[{{Rees} \& {M{\'e}sz{\'a}ros}(2005)}]{Rees&Meszaros2005}
{Rees}, M.~J. \& {M{\'e}sz{\'a}ros}, P. 2005, \apj, 628, 847

\bibitem[{{Salafia} \& {Ghirlanda}(2022)}]{Salafia&Ghirlanda2022}
{Salafia}, O.~S. \& {Ghirlanda}, G. 2022, Galaxies, 10, 93

\bibitem[{{Sari} \& {Piran}(1997)}]{Sari&Piran1997}
{Sari}, R. \& {Piran}, T. 1997, \apj, 485, 270

\bibitem[{{Shahmoradi} \& {Nemiroff}(2011)}]{Shahmoradi&Nemiroff2011}
{Shahmoradi}, A. \& {Nemiroff}, R.~J. 2011, \mnras, 411, 1843

\bibitem[{{Sobacchi} {et~al.}(2021){Sobacchi}, {Sironi}, \&
  {Beloborodov}}]{Sobacchi2021}
{Sobacchi}, E., {Sironi}, L., \& {Beloborodov}, A.~M. 2021, \mnras, 506, 38

\bibitem[{{Toffano} {et~al.}(2021){Toffano}, {Ghirlanda}, {Nava}, {Ghisellini},
  {Ravasio}, \& {Oganesyan}}]{Toffano2021}
{Toffano}, M., {Ghirlanda}, G., {Nava}, L., {et~al.} 2021, \aap, 652, A123

\bibitem[{{Tsvetkova} {et~al.}(2017){Tsvetkova}, {Frederiks}, {Golenetskii},
  {Lysenko}, {Oleynik}, {Pal'shin}, {Svinkin}, {Ulanov}, {Cline}, {Hurley}, \&
  {Aptekar}}]{Tsvetkova2017}
{Tsvetkova}, A., {Frederiks}, D., {Golenetskii}, S., {et~al.} 2017, \apj, 850,
  161

\bibitem[{{Ursi} {et~al.}(2020){Ursi}, {Tavani}, {Frederiks}, {Romani},
  {Verrecchia}, {Marisaldi}, {Aptekar}, {Antonelli}, {Argan}, {Bulgarelli},
  {Barbiellini}, {Caraveo}, {Cardillo}, {Casentini}, {Cattaneo}, {Chen},
  {Costa}, {Donnarumma}, {Evangelista}, {Feroci}, {Ferrari}, {Fuschino},
  {Galli}, {Giuliani}, {Labanti}, {Lazzarotto}, {Longo}, {Lucarelli},
  {Morselli}, {Paoletti}, {Parmiggiani}, {Piano}, {Pilia}, {Pittori},
  {Svinkin}, {Trois}, {Tsvetkova}, {Vercellone}, \& {Vittorini}}]{Ursi2020}
{Ursi}, A., {Tavani}, M., {Frederiks}, D.~D., {et~al.} 2020, \apj, 904, 133

\bibitem[{{von Kienlin} {et~al.}(2020){von Kienlin}, {Meegan}, {Paciesas},
  {Bhat}, {Bissaldi}, {Briggs}, {Burns}, {Cleveland}, {Gibby}, {Giles},
  {Goldstein}, {Hamburg}, {Hui}, {Kocevski}, {Mailyan}, {Malacaria},
  {Poolakkil}, {Preece}, {Roberts}, {Veres}, \&
  {Wilson-Hodge}}]{vonKienlin2020}
{von Kienlin}, A., {Meegan}, C.~A., {Paciesas}, W.~S., {et~al.} 2020, \apj,
  893, 46

\bibitem[{{Wang} {et~al.}(2024){Wang}, {Xie}, {Gao}, {Xiao}, {Dong}, {Zhang},
  \& {Zhi}}]{Wang2024}
{Wang}, W.-K., {Xie}, W., {Gao}, Z.-F., {et~al.} 2024, Research in Astronomy
  and Astrophysics, 24, 025006

\bibitem[{{Woosley}(1993)}]{Woosley1993}
{Woosley}, S.~E. 1993, \apj, 405, 273

\bibitem[{{Woosley} \& {Bloom}(2006)}]{Woosley2006}
{Woosley}, S.~E. \& {Bloom}, J.~S. 2006, \araa, 44, 507

\bibitem[{{Yonetoku} {et~al.}(2004){Yonetoku}, {Murakami}, {Nakamura},
  {Yamazaki}, {Inoue}, \& {Ioka}}]{Yonetoku2004}
{Yonetoku}, D., {Murakami}, T., {Nakamura}, T., {et~al.} 2004, \apj, 609, 935

\bibitem[{{Yonetoku} {et~al.}(2010){Yonetoku}, {Murakami}, {Tsutsui},
  {Nakamura}, {Morihara}, \& {Takahashi}}]{Yonetoku2010}
{Yonetoku}, D., {Murakami}, T., {Tsutsui}, R., {et~al.} 2010, \pasj, 62, 1495

\bibitem[{{Zhang} \& {Yan}(2011)}]{Zhang&Yan2011}
{Zhang}, B. \& {Yan}, H. 2011, \apj, 726, 90

\bibitem[{{Zhang} {et~al.}(2009){Zhang}, {Zhang}, {Virgili}, {Liang}, {Kann},
  {Wu}, {Proga}, {Lv}, {Toma}, {M{\'e}sz{\'a}ros}, {Burrows}, {Roming}, \&
  {Gehrels}}]{Zhang2009}
{Zhang}, B., {Zhang}, B.-B., {Virgili}, F.~J., {et~al.} 2009, \apj, 703, 1696

\bibitem[{{Zhao} {et~al.}(2014){Zhao}, {Li}, {Liu}, {Zhang}, {Bai}, \&
  {M{\'e}sz{\'a}ros}}]{Zhao2014}
{Zhao}, X., {Li}, Z., {Liu}, X., {et~al.} 2014, \apj, 780, 12

\end{thebibliography}

\end{document}